\documentclass{elsarticle}	

\usepackage{lineno,hyperref}
\modulolinenumbers[5]

\journal{Journal of Biomedical Signal Processing and Control}

\usepackage{amssymb}
\usepackage{scalerel}
\usepackage{stackengine,wasysym}

\begin{document}

\begin{frontmatter}

\title{Classification of ALS patients based on acoustic analysis of sustained vowel phonations}

\author{M. Vashkevich\fnref{myfootnote1}}
\fntext[myfootnote1]{Department of Computer Engineering, Belarusian State University of Informatics and Radioelectronics, 6 P.Brovky str., 220013, Minsk, Belarus}

\author{Yu. Rushkevich\fnref{myfootnote2}}
\fntext[myfootnote2]{Republican Research and Clinical Center of Neurology and Neurosurgery, 24 F. Skoriny Skoriny str., 220114, Minsk, Belarus}




\begin{abstract}
Amyotrophic lateral sclerosis (ALS) is incurable neurological disorder with rapidly progressive course. Common early symptoms of ALS are difficulty in swallowing and speech. However, early acoustic manifestation of speech and voice symptoms is very variable, that making their detection very challenging, both by human specialists and automatic systems. This study presents an approach to voice assessment for automatic system that separates healthy people from patients with ALS. In particular, this work focus on analysing of sustain phonation of vowels /a/ and /i/  to perform automatic classification of ALS patients. A wide range of acoustic features such as MFCC, formants, jitter, shimmer, vibrato, PPE, GNE, HNR, etc. were analysed. We also proposed a new set of acoustic features for characterizing harmonic structure of the vowels. Calculation of these features is based on pitch synchronized voice analysis. A linear discriminant analysis (LDA) was used to classify the phonation produced by patients with ALS and those by healthy individuals. Several algorithms of feature selection were tested to find optimal feature subset for LDA model. The study's experiments show that the most successful LDA model based on 32 features picked out by LASSO feature selection algorithm attains 99.7\% accuracy with 99.3\% sensitivity and 99.9\% specificity. Among the classifiers with a small number of features, we can highlight LDA model with 5 features, which has 89.0\% accuracy (87.5\% sensitivity and 90.4\% specificity).

\end{abstract}


\end{frontmatter}


\section{Introduction}
Amyotrophic lateral sclerosis (ALS) is a fatal neurodegenerative disease involving the upper and lower motor neurons. There are two main forms of ALS which differ by onset: spinal form (first symptoms manifest in the arms and legs) and bulbar form (voice and/or swallowing difficulties are often the first symptoms). Progressive bulbar motor impairment due to ALS leads to deterioration in speech and swallowing function~\cite{Green-2013}. The abnormalities in speech production, phonation and articulation due to neurological disorders is referred to as \textit{dysarthria}. Dysarthria develops in more than 80\% of affected by ALS individuals at some point during the disease's course~\cite{Duffy-2013}.
Currently the diagnosis of ALS is based on clinical observations of upper and lower motor neuron damage in the absence of other causes. Due to the lack of clinical diagnostic markers of ALS, the pathway to correct diagnosis in average takes 12 months~\cite{Iwasaki-2001}.

During the last years objective evaluation of voice and speech has gained popularity as a means of detecting early signs of neurological diseases~\cite{Benba-16,Rusz-2011,Orozco-16}. It can be explained by the fact that speech is accomplished through complex articulatory movements, requires precise coordination and timing and therefore is very sensitive to violations in the peripheral or central nervous system~\cite{GomezVilda-2015,Guerra-03}. 
Recent studies suggested that acoustic voice and speech analysis might provide useful biomarkers for diagnosis and remote monitoring of ALS patients~\cite{Norel-18,Spangler-17}. The advantage of using voice/speech signals is the capability of using smartphone or tablet for recording patients at home conditions without the logistical difficulty in a clinical environment~\cite{Benba-16,Kwanghoon-2018}.

The main goal of this work is automatic detection of ALS patients with or without bulbar disorders (i.e. classification of healthy controls vs. patients with ALS) based on sustained vowel phonation (SVP) test.
Our long-term aim is to build automated system for classification of neuromotor degenerative disorders based on analysis of SVP test. Therefore, we consider the problem of binary classification of the voice recording to be belonging to ALS patient or healthy person as a first step toward this aim. We chose sustained vowel phonation test among different diagnostic speech tasks due to its simplicity and wide spreading in medical practice. Besides all, recent research shows~\cite{Tsanas-2012} that using SVP test it is possible to detect persons with Parkinson’s disease. This give us hope that this test can be effective for ALS detection.

Sustained phonation is a common speech task used to evaluate the health of the phonatory speech subsystem~\cite{Rusz-2011}. By using SVP test the following characteristics of voice can be assessed: pitch, loudness, resonance, stain, breathiness, hoarseness, roughness, tremor,  etc~\cite{Benba-16,Guerra-03,Gomez-2019}. However, it can be argued that some of the vocal abnormalities in continuous speech might not be captured by use of sustained vowels, but the analysis of continuous speech is much more complex due to articulatory and other linguistic confounds~\cite{Gomez-2019}. One more argument is that the use of sustained vowels is commonplace in clinical practice~\cite{Baken-2000}. Besides all this, early study~\cite{Silbergleit-97} had been reported that abnormal acoustic parameters of the voice were demonstrated in ALS subjects with perceptually normal vocal quality on sustained phonation. Also in~\cite{Graaff-2009} it was reported that glottic narrowing due to vocal cord dysfunction (that can be assessed using SVP test) is one of ALS symptoms. 

SVP test is widely used for detecting and diagnosing of different neurological diseases such as Parkinson's, Alzheimer, Dystonia and others~\cite{Benba-16,Rusz-2011}. For example,  it has been shown in~\cite{Tsanas-2012} that classifier based on the features extracted form SVP test allows one to discriminate Parkinson's disease subjects from healthy controls (HC) with almost 99\% overall classification accuracy.  However, there are few studies dedicated to the detection of the ALS based on SVP test. In~\cite{Guerra-03} SVP was used along with the other speech tests for dysarthria classification. 
Sustained phonation also was used for assessing laryngeal subsystem within a comprehensive speech assessment battery in~\cite{Yunusova-13}. But in the majority of prior works {\it running speech} test that consist in reading of specially-designed passage was used for ALS detection~\cite{,Norel-18,Kwanghoon-2018,Mujumdar-10,Illa-18}. In~\cite{Spangler-17} rapid repetition of syllable (pa/ta/ka), which is often referred to as {\it diadochokinetic task} (DDK) was used for automatic ALS detection. Some studies use kinematic sensors to model articulation for ALS detection~\cite{Bandini-17}, however this approach is invasive in nature and less attractive compared to non-invasive speech test.

The purpose of this work is to investigate the possibility of designing a classifier for detection of patients with ALS based on the sustained phonation test. Traditionally, vowel /a/ is used in SVP test, however, in our study along with /a/ we have used vowel /i/. This decision is based on preliminary results of works~\cite{Lee-2017,Vashkevich-18,Gvozdovich-19}, that provide evidence that information contained in these vowels might allow to obtain classifier with high performance. This work is based on the analysis of the sustained phonation of vowels /a/ and /i/, in contrast to other studies that extract vowels from running speech tests (see e.g.~\cite{Vashkevich-18,Gvozdovich-19}). 

The remainder of the paper is organized as follows; Section~\ref{sec:analysis} provides information about methods of acoustic analysis used for feature extraction. The voice data used in this study along with various methods of feature selection, classification and validation are presented in section~\ref{sec:experiments}. In section~\ref{sec:results} we present the results of our findings and discuss the interpretation of them. Section~\ref{sec:conclusion} provides conclusion on the work.

\section{Acoustic analysis}
\label{sec:analysis}
Bulbar system that is affected by ALS is considered as a part of the larger speech production network and comprises of four distinct subsystems~\cite{Green-2013}: respiratory, phonatory, articulatory, and resonatory. In this short review of acoustic features, we indicate which subsystem is described by each feature. 

\subsection{Perturbation measurements}

\subsubsection{Jitter} Jitter (i.e. frequency/period perturbation) is the measure of variability of fundamental period from one cycle to the next. As far as jitter estimates short-term variations it can not be accounted to voluntary changes in F0. Therefore jitter is intended to provide an index of the stability of the phonatory subsystem. 
High level of jitter results from diminished neuromotor and aerodynamic control~\cite{Baken-2000}. The jitter has been used as an indicator of the voice quality that characterizes the severity of dysphonia~\cite{Castro-13}. In this study we have used following popular jitter measures~\cite{Kasuya-82}: 

1) local jitter ($J_{loc}$) that is defined as average difference between consecutive periods, divided by the average period:
\begin{equation}
  J_{loc} = \frac{\textstyle \frac{1}{N-1}\sum\limits_{i=1}^{N-1}|T_0(i)-T_0(i+1)|}{
  \textstyle \frac{1}{N}\sum\limits_{i=1}^{N}T_0(i)}\times 100,
  \label{eq:J_loc}
\end{equation}
where $T_0(i)$ is the duration of $i$-th fundamental period in seconds, $N$ is the number of extracted periods; 


2) period perturbation quotient ($J_{ppq}$) to quantify the variability of pitch period evaluated in $L$ consecutive cycles: 

\begin{equation}
  J_{ppqL} = \frac{\displaystyle  \frac{1}{N-L+1}\sum_{i=1+\frac{L-1}{2}}^{N-\frac{L-1}{2}}\Bigl|T_0(i)-\frac{1}{L}\sum_{k=i-\frac{L-1}{2}}^{i+\frac{L-1}{2}}T_0(k)\Bigr|}{
  \displaystyle \frac{1}{N}\sum_{i=1}^{N}T_0(i)}\times 100.
  \label{eq:J_ppq}
\end{equation}
In this work, we used the parameter $L$ equal to 3, 5 and 55.

\subsubsection{Shimmer} 
Shimmer is an amplitude perturbation measure that characterize the extent of variation of expiratory flow during the phonation. This feature can be considered as characteristic of the respiratory subsystem. Basic shimmer measure ($S_{loc}$) is defined as average absolute difference between the amplitude of consecutive periods, divided by the average amplitude:

\begin{equation}
  S_{loc} = \frac{ \frac{1}{N-1}\sum\limits_{i=1}^{N-1}|A(i)-A(i+1)|}{
   \frac{1}{N}\sum\limits_{i=1}^{N}A(i)}\times 100,
  \label{eq:S_loc}
\end{equation}
where $A(i)$ is the amplitude of the $i$-th pitch period.

 $S_{loc}$ fall under influence of long-term changes in vocal intensity~\cite{Baken-2000}. To eliminate the effects of amplitude ``drift'' and get a truer index of underlying shimmer it has been suggested to measure amplitude perturbation quotient (APQ)~\cite{Kasuya-82}. APQ quantify whether the  amplitude of pitch period duration is smooth over $L$ consecutive cycles:
 \begin{equation}
   S_{apqL} = \frac{ \frac{1}{N-L+1}\sum\limits_{ i=1+\frac{L-1}{2}}^{ N-\frac{L-1}{2}}\Bigl|A(i)-\frac{1}{L}\sum\limits_{ k=i-\frac{L-1}{2}}^{  i+\frac{L-1}{2}}A(k)\Bigr|}{
   \frac{1}{N}\sum_{i=1}^{N}A(i)}\times 100,
   \label{eq:S_apq}
 \end{equation}
Typically the parameter $L$ takes value 3, 5, 11 or 55~\cite{Rusz-2011,Orozco-16,moran-2006}. We used all of those options in our study.

\subsubsection{Directional perturbation factor} 
Directional perturbation factor (DFP) is a measure of perturbation that ignores the magnitude of period perturbation: it depends on the number of times that frequency changes shift direction~\cite{Baken-2000}. The DFP calculation consists of two steps. At the first step the difference between adjacent fundamental periods is calculated:
\begin{equation} 
\Delta T_0(i) = T_0(i) - T_0(i-1).
\end{equation}
At the second step the number of sign changes in sequence of $\Delta T_0(i)$ is calculated:
$$ N_{\Delta\pm} = \frac12 \sum_{i=2}^{N} |\mathrm{sign}(\Delta T_0(i)) - \mathrm{sign}(\Delta T_0(i-1))|.$$
Finally, DFP parameter is obtained as follows:
 \begin{equation}
\mathrm{DFP} = \frac{N_{\Delta\pm}}{N}\times 100,
\end{equation}
where $N$ is the total number of fundamental periods. 

\subsection{Noise measurements}
The existence of noise energy, broadly understood as that outside of harmonic components during sustained phonation, is the result of incomplete closure of the vocal folds during the phonation, indicative of an interruption of the morphology of the larynx~\cite{moran-2006}. We used two different noise measurements: harmonic-to-noise ratio (HNR)~\cite{Boersma-1993} and glottal-to noise excitation ratio (GNE).

\subsubsection{HNR}
The HNR measures the ratio between periodic (or harmonic) component and non periodic (or noise) component of the voice signal. Sonorant and harmonic voices are characterized by high HNR values. A low HNR denotes that voice comprise increased amount of noise. For calculation of HNR we used mathematical background presented in~\cite{Boersma-1993}. At the beginning, for a voice signal a normalized autocorrelation function $AC_V(\tau)$ is calculated. Then, the first local maximum outside 0 (with corresponded lag $\tau_{max}$) is searched. The normalized autocorrelation $AC_V(\tau_{max})$ represents the relative power of the periodic component of the signal (while full power $AC_V(0)=1$). Finally, HNR is calculated as
 \begin{equation}
\mathrm{HNR} = 10\log_{10} \frac{ AC_V(\tau_{max})}{1-AC_V(\tau_{max})}.
\end{equation}

\subsubsection{GNE} 
GNE measures the amount of excitation in voice due to the vibration of the vocal folds relative to the excitation noise due to the turbulence in the vocal tract~\cite{Orozco-15}. The GNE is often associated with the breathiness~\cite{Guerra-03,Awan-2009} and therefore can be considered as characteristics of phonatory subsystem.

Calculation of the GNE is based on the correlation between Hilbert envelopes of three different frequency channels~\cite{Michaelis-1997}. Since full band signal simultaneously excited by a single glottis closure the envelopes in all channels have the same shape. This leads to high correlation between envelopes. However, in case of turbulent signals a narrowband noise is excited in each frequency channel. These narrow band noise signals are uncorrelated. Thus, interband correlation can be used to measure the amount of turbulence in a signal.

Calculation of the GNE factor is consist in the following steps:
\begin{enumerate}
\item	Down sampling the signal to 8 kHz.
\item   Divide signal into 30 ms overlapping frames with 10 ms hop size. For each frame execute steps 3--7.
\item	Inverse filtering of the signal by calculating the linear prediction error signal, using a predictor of 10-th order estimated by the autocorrelation method~\cite{Huang-2001} with Hamming window.
\item	Calculating the Hilbert envelopes of three different frequency bands with 1000 Hz bandwidth and central frequencies at 500, 1500 and 2500 Hz.
\item	Calculating the cross correlation function between every pair of envelopes for which the difference of their center frequency is equal or greater than half the bandwidth.
\item	Pick the maximum of each correlation function.
\item	The GNE for the current frame is equal to the maximum of the maximums obtained in step 6.
\item   Compute the mean value of GNE and its standard deviation.
\end{enumerate}

\subsection{Spectral parameters}

\subsubsection{MFCC} Mel-Frequency Cepstral Coefficients (MFCCs) is the most widely used feature in speech-related applications such as speaker identification and recognition. 
Moreover, recent studies have shown promising results on the identification of voice pathology with MFCCs ~\cite{ Benba-16, Spangler-17,Tsanas-2012,Godino-2006}.
%
MFCCs can detect subtle changes in the motion of the articulators (tongue, lips), which are known to be affected in many neurological diseases~\cite{Tsanas-2012}. They have been used for detecting of hypernasality due to the velopharyngeal insufficiency in~\cite{Dubey-2018}. In~\cite{Orozco-15} the usage of MFCCs is argued by its ability of modelling changes in the speech spectrum, especially around the first two formants (F1 and F2), where most of the energy of the signal is concentrated. The work~\cite{Godino-2006} showed that MFCCs have an inherent ability to model an irregular movement of the vocal folds, or a lack of closure due to a change in the properties of the tissue covering vocal folds. Therefore MFCCs can be considered as parameters describing both resonatory and articulatory subsystems. 

MFCC parameters~\cite{Godino-2006,Huang-2001} are obtained by applying discrete cosine transform over the logarithm of the energy calculated in several mel-frequency bands:
 \begin{equation}
\mathrm{MFCC}(m)  = \sum_{k=1}^{M} \ln{S(k)} \cos \bigl[ m (k-0.5) \frac{\pi}{M}\bigr] .
\end{equation}
where $M$ is the number of uniform frequency bands in the mel scale, $m=1,2,\dots, L$, and $L$ is the order of the MFCC coefficients. The energy of frequency bands are calculated using $N$-point magnitude spectrum $X(j)$ of the frame of the voice signal:
 \begin{equation}
S(k)  = \sum_{j=1}^{N} W_k(j) X(j), \quad k=1,2,\dots,M,
\label{eq:S_k}
\end{equation}
where $W_k(j)$ is the triangular weighting function~\cite{Huang-2001} associated with $k$-th band.

In our study we used $L=12$ MFCC parameters that computed within windows of 25-ms length and 10-ms time shift. Magnitude spectrum $X(j)$ is calculated in the range $[0; 4000]$ Hz and averaged within $M=20$ uniform mel-frequency bands (see (\ref{eq:S_k})). The first ($\Delta$) derivatives of MFCC have also been calculated since they provide information about the dynamics of the time-variation in MFCC parameters. A priori, these features can be considered as significant because the disorder lowers stability of the voice signal~\cite{Godino-2006}; therefore lager time-variations of the parameters may be expected in ALS voice relative to normal voice.

Because the MFCCs are a timeseries, we averaged the MFCCs across the time domain to consolidate them to a single set of coefficients. Finally 12 MFCCs and 12 $\Delta$MFCCs are evaluated for each voice recording.

\subsubsection{Formants} 
Changes of formant frequencies during the vowel phonation due to dysarthria have been reported in many studies~\cite{Lee-2017,Kent-99,Tomik-2010,Turner-95,Weismer-92}.
The most frequently reported abnormalities of vowel production include: centralization of formant frequencies~\cite{Lansford-2014}, reduction of the vowels space area~\cite{Turner-95}, and abnormal formant frequencies for high vowels and front vowels~\cite{Vashkevich-18,Kent-99}. In~\cite{Weismer-92} it was shown that in patients with ALS measurement of the F2 slope (or F2 transition) is correlated with overall speech intelligibility score. Also features derived from statistics of the first (F1) and second (F2) formant frequencies (and their trajectories) have shown good performance for predicting speaking rate decline in ALS~\cite{Horwitz-16}. Though SVP test cannot reflect the dynamics of formant frequency trajectories, we still can use the values of formant frequencies as source of information. In~\cite{Gvozdovich-19,Lee-2019} it was shown that the value of F2 for vowel /i/ appears to be a good feature for discriminating between patients with ALS and healthy control group. In this study we use second formant of vowel /i/ ($F2_i$) and Euclidean distance (convergence) between the vowels /i/ and /a/:
\begin{equation}
F2_{conv}=\bigl| F2_i - F2_a \bigr|.
\end{equation}
Study~\cite{Lee-2017} have shown that convergence of the F2 of vowels /i/ and /a/ is much stronger in speakers with dysarthria due to ALS, than in healthy speakers. Both features ($F2_i$ and $F2_{conv}$) are prove to be a highly informative for ALS detection using running speech test~\cite{Gvozdovich-19}.

\subsubsection{Distance between the spectral envelopes of the vowels}
\begin{figure}[th]
    \centering
    \begin{minipage}{0.49\textwidth}
        \centering
        \includegraphics[width=0.99\textwidth]{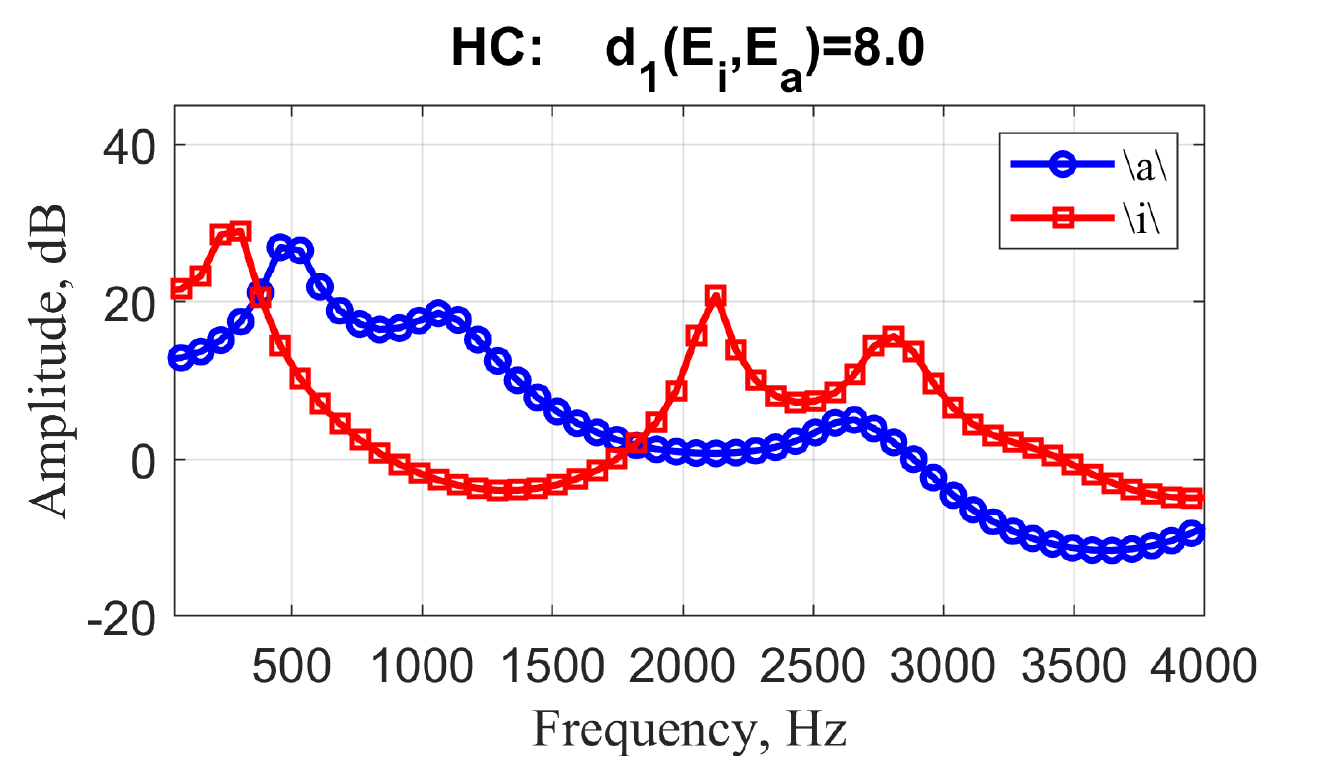} 
       \small (a)
    \end{minipage}\hfill
    \begin{minipage}{0.49\textwidth}
        \centering
        \includegraphics[width=0.99\textwidth]{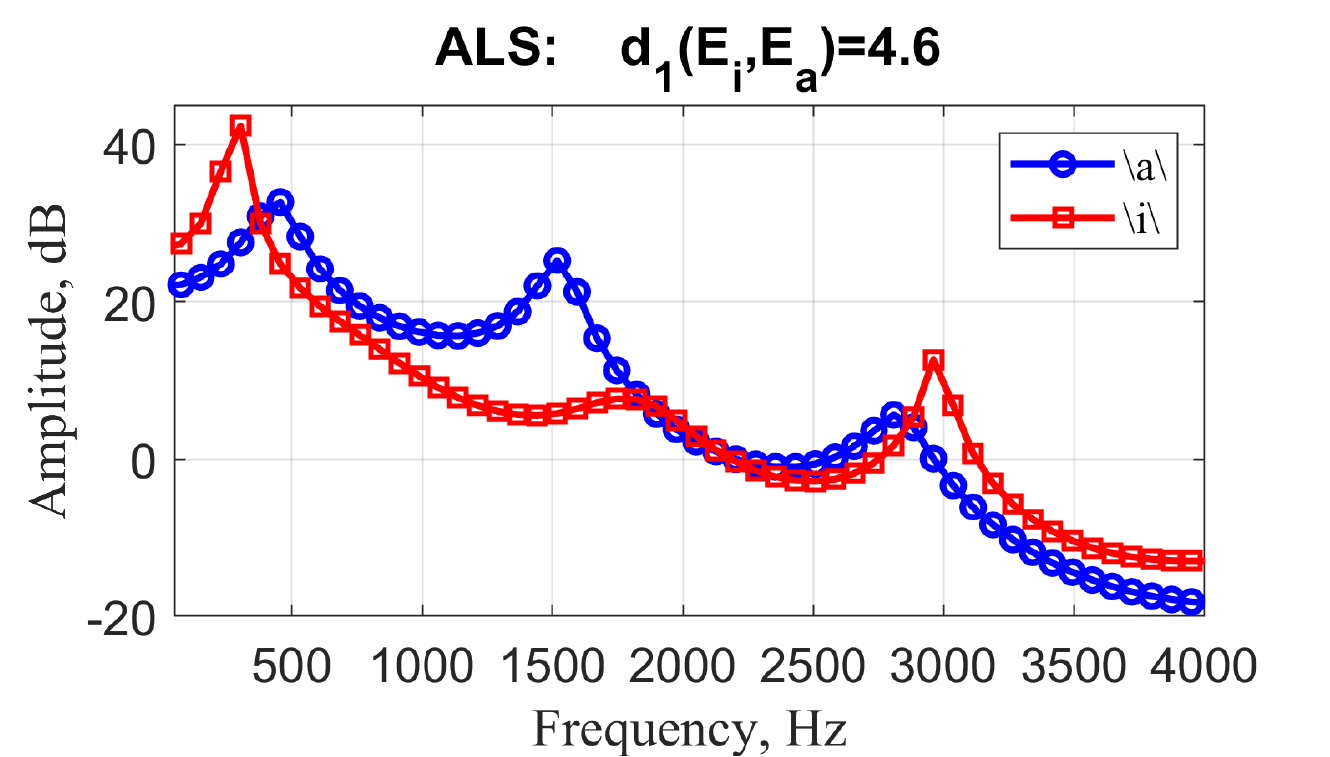} 
       \small (b)
    \end{minipage}
\caption{Envelopes of vowels /a/ and /i/: (a) healthy speaker; (b) ALS patient}
\label{fig:envs}    
\end{figure}

In~\cite{Vashkevich-18} it was suggested to use distance between the spectral envelopes of the vowels /a/ and /i/ to quantify the amount of articulatory undershoot. The joint analysis of envelopes of vowels /a/ and /i/ of persons with ALS have revealed an increased similarity between the shapes of these envelopes. The similarity between the envelopes is assessed using $l_1$-norm distance metric 
\begin{equation}
d_1(E_i,E_a) =\frac{1}{P} \sum_{k=1}^{P}|E_i(k) - E_a(k)|,
\label{eq:d1}
\end{equation}
where $E_i(k)$ is envelope of the vowel /i/, $E_a(k)$ is envelope of the vowel /a/, $P$ is the number of points in frequency domain. The spectral envelopes of the vowels were estimated using all-pole modelling~\cite{Huang-2001}. An example of vowel envelopes from healthy individual are shown in figure~\ref{fig:envs},a. A typical example of envelopes with a high degree of similarity is given in figure~\ref{fig:envs},b. 

\subsection{$F0$ contour based parameters}
\subsubsection{Phonatory frequency range} 
Phonatory frequency range (PFR) is defined as semitone difference between lowest ($F0_{\mathrm{low}}$) and highest ($F0_{\mathrm{high}}$) fundamental frequencies~\cite{moran-2006}:
 \begin{equation}
\mathrm{PFR} = 12\frac{\log_{10}\bigl(F0_{\mathrm{high}}/F0_{\mathrm{low}}\bigr)}{\log_{10}2}.
\end{equation}
This parameter measures the degree of variability in fundamental frequency contour and characterizes the functioning of the phonatory subsystem.

\subsubsection{Pitch period entropy}
Pitch period entropy (PPE) is a highly informative feature proposed in~\cite{Little-2008}  to assess the degree of loss of control over the stationary voice pitch during sustain phonatition (due to Parkinson's disease). We have used this measure in our study since the ALS also affects the ability to control the stability of voice pitch.

The calculation of PPE is based on the following observations: 1)  the healthy voice has natural pitch variation characterized by smooth vibrato or microtremor~\cite{Baken-2000, Little-2008}; and 2) speakers with naturaly high-pitch voices have much lager vibrato and microtremor than speakers with low-pitch voices. PPE measurement takes into account both these factors. The natural smooth variations is removed prior to measuring the extent of such variations (first factor) and pitch transformation to perceptually-relevant, logarithmic semitone scale is applied (second factor). The algorithm of PPE calculation used in this study is given below.

\begin{figure}[thb]
  \centering
  \includegraphics[width=125mm]{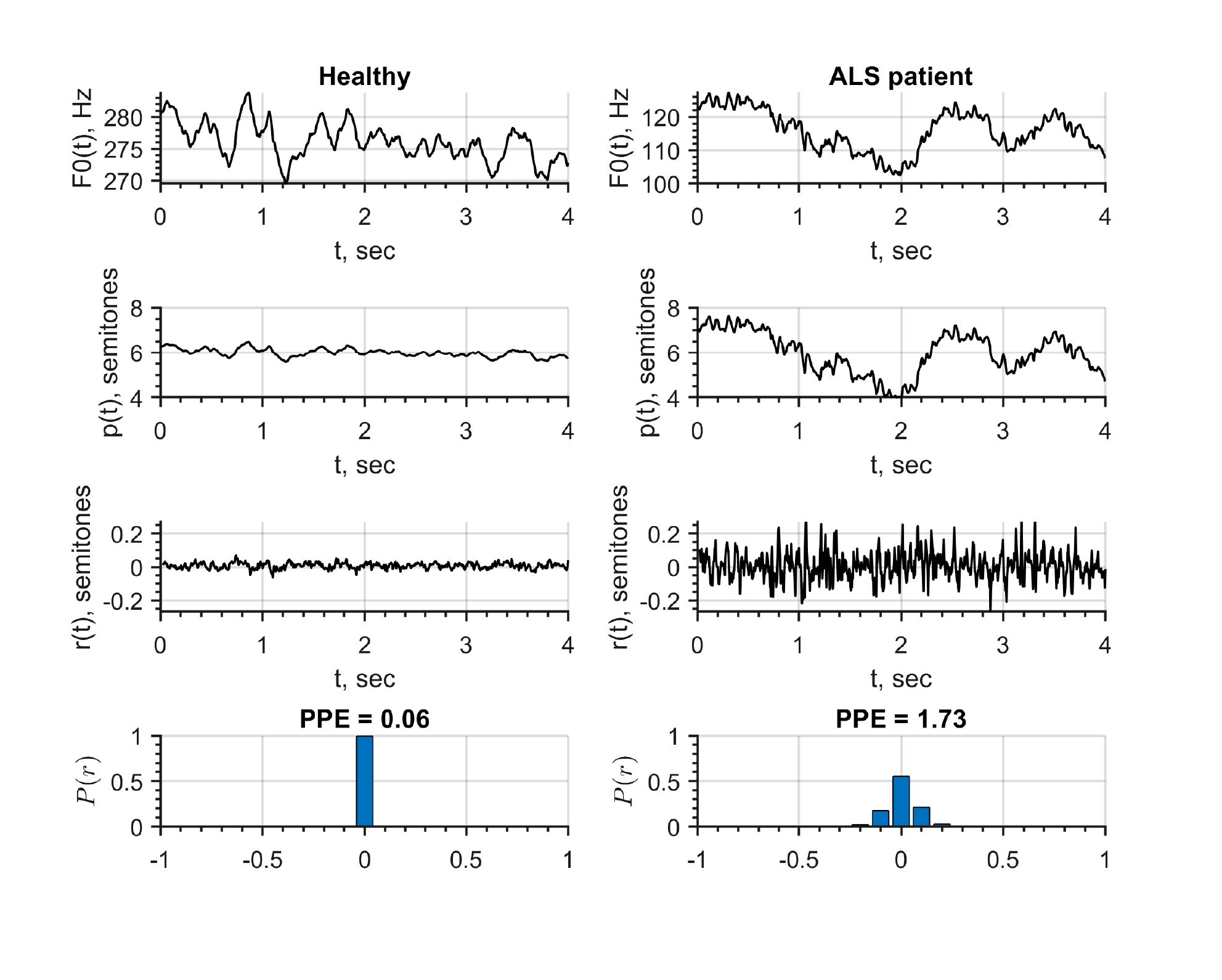}
  \caption{Details of PPE calculation, left column: healthy subject, right column: ALS patient. Rows from the top: extracted F0, pitch $p(t)$ in semitone scale, residual signal $r(t)$ after spectral whitening filter, probability densities $P(r)$ of residual pitch period $r$}
  \label{fig:ppe_steps}
\end{figure}

\begin{enumerate}
\item Estimation of $F0(m)$ contour with 5~ms time step using IRAPT algorithm~\cite{Azarov-12};
\item Transformation of  $F0(m)$ contour to semitone scale: 
\begin{equation}
p(m) = 12 \frac{\log_{10}\bigl(F0(m) /f_{\mathrm{low}}\bigr)}{\log_{10}2},
\end{equation}
 where $f_{\mathrm{low}}$  is lower octave band limit, calculated considering that mean value of pitch  correspond to center of this octave: $$f_{\mathrm{low}} = \mathrm{mean}(F0)/\sqrt{2}.$$ 
\item Applying whitening filter to  $p(m)$ signal to remove healthy, smooth variation: 
\begin{equation}
r(m)  = \sum_{i=0}^{M}a_i p(m-i), \quad a_0=1,
\end{equation}
where $a_i$ is linear prediction coefficients (LPC) estimated using covariance method~\cite{Huang-2001}, $M$ is the predictor order. We used $M=2$;
\item Calculation of discrete probability distribution of occurrence of relative semitone variations $P(r)$ by computing normalized histogram in $N=31$ equal-sized bins $r_i$ ($i=1,2,\dots,N$) in the range form $-1.5$ to $1.5$;
\item Calculation the entropy distribution $P(r)$ obtained on previous step:
\begin{equation}
  \mathrm{PPE} =- \sum_{i=1}^{N} P(r_i) \log_2P(r_i),
\end{equation}
\end{enumerate}
The larger the measure of entropy, the more the observed variations exceed the natural level of variation of the fundamental frequency in a healthy voice. The fig.~\ref{fig:ppe_steps} give an example that illustrates the process of calculation of  PPE measure.

 Figure~\ref{fig:ppe_steps} shows that semitone pitch sequence $p(t)$ of healthy voice is quite stable and has signs of small regular vibrato. After eliminating this healthy vibrato with whitening filter, the distribution of residuals $r(t)$ shows strong peak at zero. This leads to small value of entropy. On the contrary, for ALS voice the semitone pitch sequence has significant irregular variation,  the distribution of residuals is spread over a wider range as a result the larger value of entropy is obtained.

\subsubsection{Tremor (vibrato) analysis}
\begin{figure}[th]
    \centering
    \begin{minipage}{0.49\textwidth}
        \centering
        \includegraphics[width=0.99\textwidth]{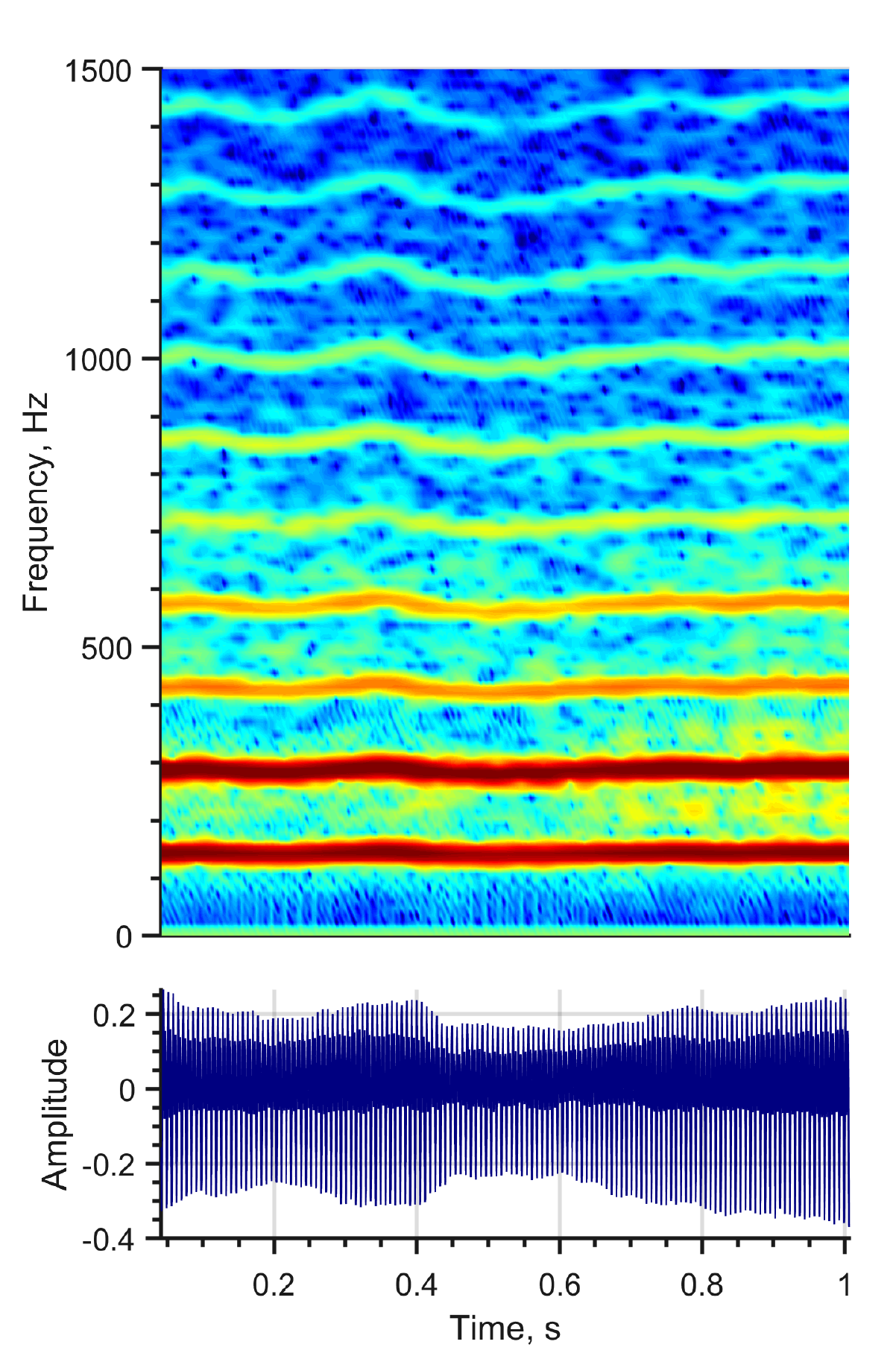} 
       \small (a)
    \end{minipage}\hfill
    \begin{minipage}{0.49\textwidth}
        \centering
        \includegraphics[width=0.99\textwidth]{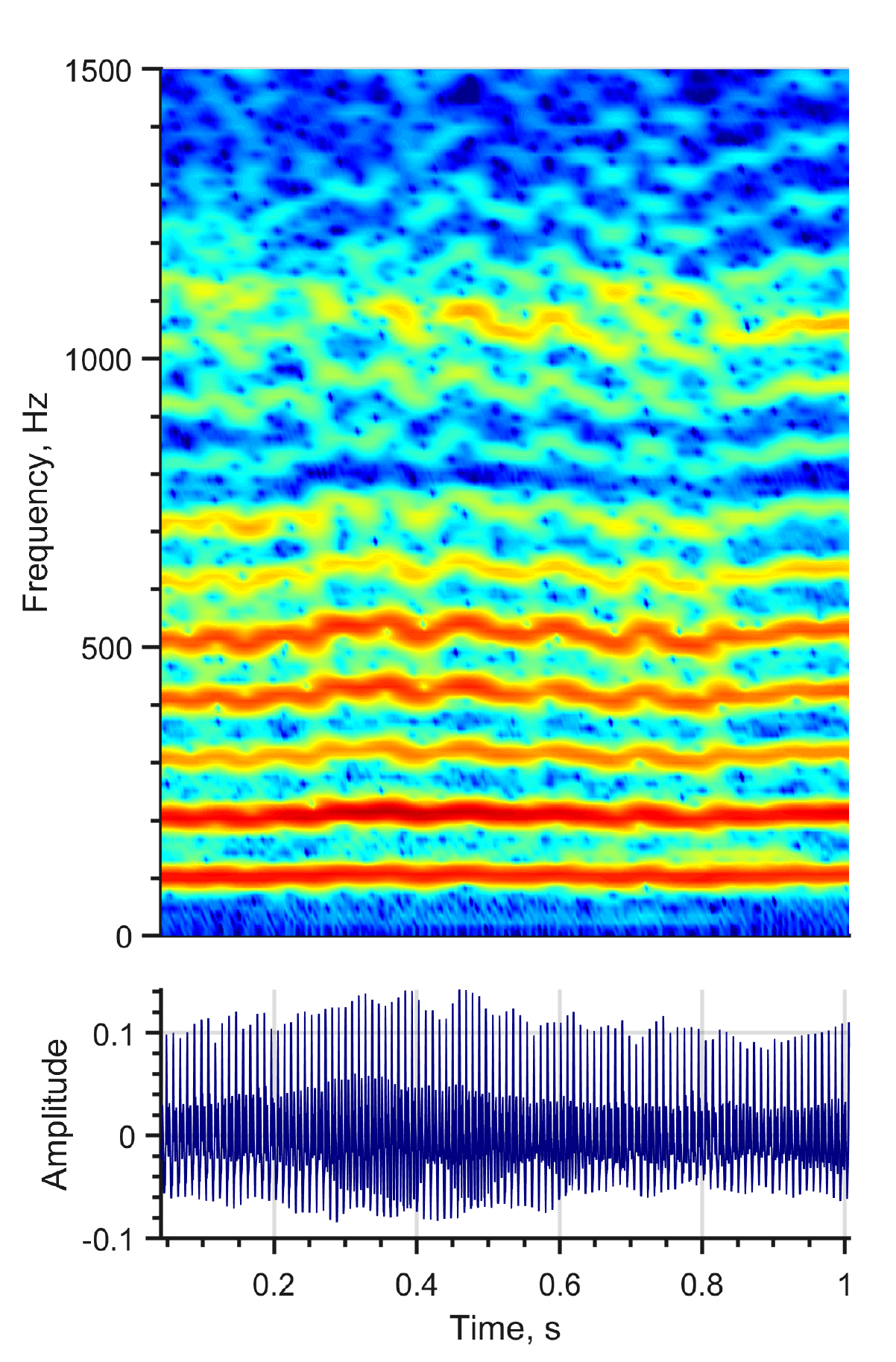} 
       \small (b)
    \end{minipage}
\caption{Time-frequency representation of vowel phonation /a/: (a) speaker from HC group (men, 60 years old); (b) ALS patient (men, 67 years old, subject code in voice base 039)}
\label{fig:vib}    
\end{figure}

Vocal tremor is involuntary quasi-sinusoidal modulation in energy and F0 contour appeared during sustained phonation~\cite{Peplinski-2019}. In our study we consider only the modulation in F0 contour. Some authors distinguish wow (oscillation of 1-2 Hz), tremor (oscillation of 2-10 Hz) and flutter (oscillation of 10-20 Hz)~\cite{Kent-99}. An example  of vowel phonation for a patient with a rapid tremor (or flutter) is given in figure~\ref{fig:vib},b (the voice is taken from the database used in experiments). 

An essential distortion can be seen when compared spectrogram of a ALS patient  (figure~\ref{fig:vib},b) with spectrogram of a normal subject (figure~\ref{fig:vib},a). In particular, in figure~\ref{fig:vib} narrowband spectrograms are shown (long 84 ms analysis window have been used for their calculation). Thus it can be seen substantial changes in harmonics behaviour. Normal voice shows stable harmonics with low variation, while harmonics of pathological voice  exhibiting high frequency quasi-sinusoidal modulations.

In~\cite{Peplinski-2019} in order to characterize the tremor the average spectra of F0 contour is analysed in frequency band from 3 to 25 Hz. However, as reported in~\cite{Aronson-92} the most essential frequency peaks of person with ALS lies within the range 6 to 12 Hz. It seems that sum of the amplitudes of spectral components in  frequency band $[6,\;12]$ Hz could be a good feature for detection of ALS voices. However, normal voices also have inherent modulations (some times called vibrato) in range 5 to 8 Hz~\cite{Nakano-06}. Thus vibrato frequency bands of healthy and ALS voices  are overlapped. So, for obtaining a new feature, that characterizes the extent of pathological modulations in F0 contour we decide to analyse the amplitudes of spectral components in range from 9 to 14 Hz. The obtained feature is referred to as \textit{pathological vibrato index} (PVI) and presented in~\cite{Vashkevich-19}. The algorithm for PVI calculation is given below

\begin{enumerate}
\item Estimation of $F0(m)$ contour with 5~ms time step using IRAPT algorithm~\cite{Azarov-12};
\item Normalization of F0 contour: 
\begin{equation}
 F0'(m)=\frac{F0(m)}{\mathrm{mean}(F0)};
\end{equation}
\item Bandpass filtering of $F0'(m)$ using 3-th order Butterworth IIR with pass band $[9,\;14]$ Hz; 
\item Amplitude spectrum $A_{F0}(f)$ estimation using Welch's method with windows of 1 sec length and 95\% overlap;
\item Calculation of pathological vibrato index:
\begin{equation}
  \mathrm{PVI} = \sum_{f\in [9,\; 14] \textit{ Hz}}A_{F0}(f).
  \label{eq:PVI}
\end{equation}
\end{enumerate}

\begin{figure}[thb]
  \centering
  \includegraphics[width=125mm]{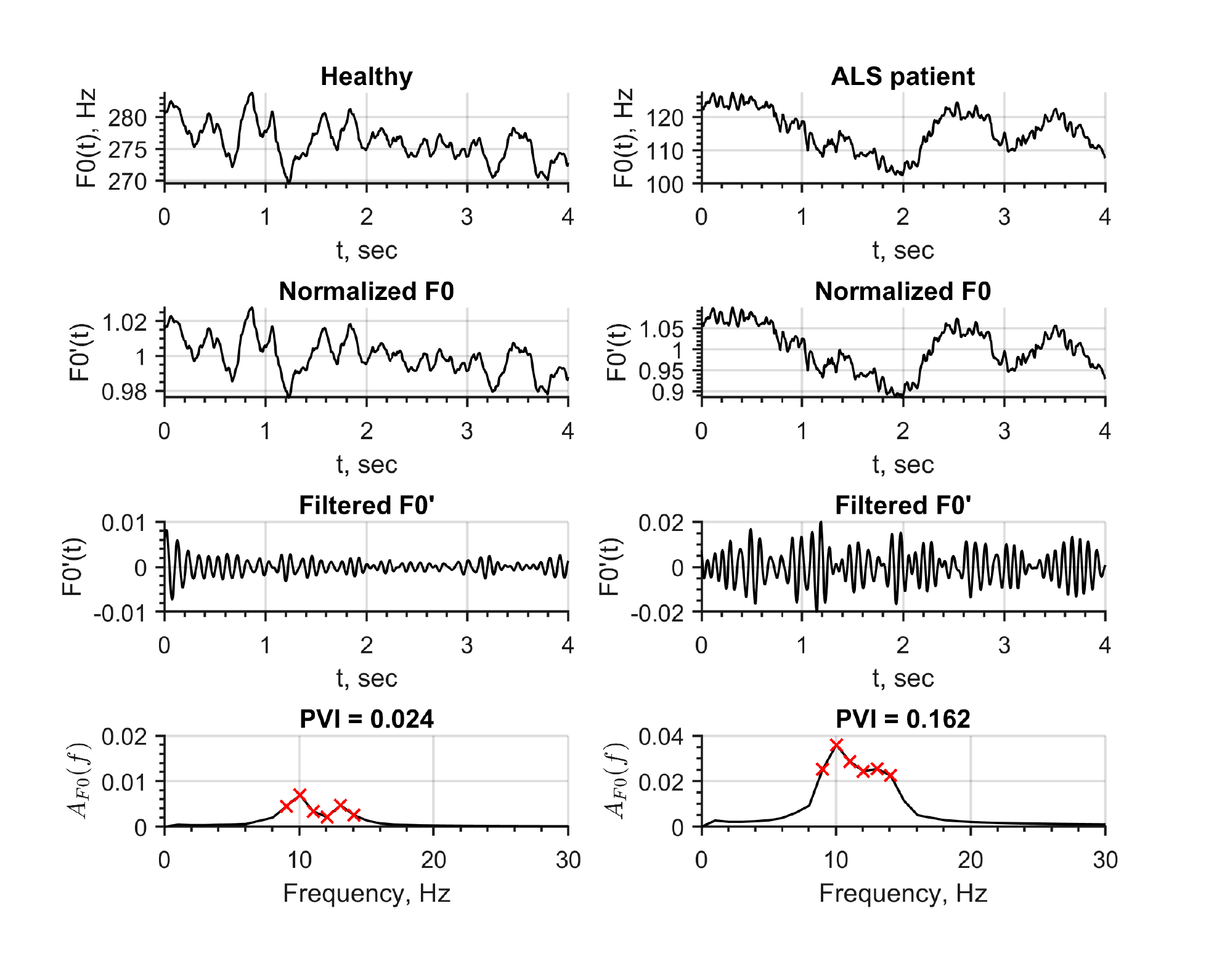}
  \caption{Left column: normal case, right column: pathological case. Rows from the top: Extracted F0, normalized F0 contour, IIR filtered F0 contour, amplitude spectrum $A_{F0}(f)$, amplitudes used for PVI calculation are indicated by red x-marks}
  \label{fig:pvi_steps}
\end{figure}

Figure~\ref{fig:pvi_steps} shows the steps of the PVI calculation for a typical normal and pathological case. It can be seen that frequency components of amplitude spectrum $A_{F0}(f)$ in the range from 9 to 14 Hz are significantly higher for the ALS voice than for a healthy voice.

\subsubsection{Analysis of the harmonic structure of the vowels}
Harmonic structure of sustained vowel has been recognized as a important and informative feature for voice pathology identification~\cite{Guerra-03,Cordeiro-2018}. Incomplete glottal closure during phonation, which allows the air to escape, is one of the factors that  makes voice more breathy. In particular, for vowel /a/ this produce a disturbance of harmonic structure:  amplitude of first harmonic (H1) becomes higher than the second (H2)~\cite{Cordeiro-2018}.

One of the important aspect of voice quality is stability of harmonics structure during the phonation process. Evaluation of harmonics structure can be considered as feature for description the excitation source (a driving force for voice production). The difficulty in estimation of harmonic parameters is that they depend on the fundamental frequency F0. In this study we have used voice analysis based on fixed number of fundamental periods (alternatively it can be considered as pitch synchronized voiced analysis). We focused on extracting mean and standard deviation (SD) of the first eight harmonics of the vowels. Given a voice signal $s(n)$ the analysis process can be summarized in the following steps.
\begin{enumerate}
	\item Split $s(n)$ into fundamental periods using waveform matching method with phase constrain~\cite{Vashkevich-19}.
	\item Divide $s(n)$ into $N_f$ overlapping frames that containing $N_c$ fundamental periods with one period overlap. For each frame $s^{(i)}(n)$, $i=1,\dots N_f$ execute steps 3--5.
	\item Interpolate $s^{(i)}(n)$ into $I\times N_c$ equidistant time points: $s^{(i)}(n)\rightarrow \widehat{s}^{(i)}(m)$.
	\item Apply Hamming window $h(m)$ to interpolated frame and compute discrete Fourier Transform (DFT): $\widehat{S}^{(i)}(k) =\mathrm{DFT}[\widehat{s}^{(i)}(m)h(m)]$. 
	\item Extract harmonic amplitudes: 
	$$ h_p(i)=\bigr|\widehat{S}^{(i)}(p\times I)\bigl|\;\;p=1,2\dots 8.$$
	\item Scale the harmonic amplitudes as $$\widetilde{H}_p(i)=20\log_{10}\biggr(\frac{h_p(i)}{\max\limits_{p\in[1,\;8],\;i\in[1,\;N_f]}\bigr\{h_p(i)\bigl\}}\biggl).$$
	\item Compute mean and SD for scaled harmonic amplitudes 
	$$ \mathrm{H}p^\mu = \mathsf{E}\{\widetilde{H}_p\},\;\; \mathrm{H}p^\sigma = \sqrt{\mathsf{E}\{\bigl(\widetilde{H}_p - \mathrm{H}p^\mu\bigr)^2\}} $$
	\item Compute additional feature -- inverse of the sum of absolute value of $\mathrm{H}p^\mu$ and $\mathrm{H}p^{\sigma}$:
	\begin{equation}
	\mathrm{RelH}p = \frac{1}{|\mathrm{H}p^\mu|+\mathrm{H}p^\sigma}.
	\label{eq:RelHp}
	\end{equation}
\end{enumerate}

The intuition behind the feature (\ref{eq:RelHp}) is that strong and stable harmonic should have low scaled amplitude $ |\mathrm{H}p^\mu|$ and low deviation $\mathrm{H}p^{\sigma}$ and therefore high value of $ \mathrm{RelH}p$.

In this study the following parameters of the procedure were used: $N_c=8$ and $I=512$.

\section{Experiments}
\label{sec:experiments}
\subsection{Database}
Voice database\footnote{The database available online at \url{https://github.com/Mak-Sim/Minsk2020_ALS_database}} used in this study was collected in Republican Research and Clinical Center of Neurology and Neurosurgery (Minsk, Belarus). It consists of 128 sustained vowel phonations (64 of vowel /a/ and 64 of vowel /i/) from 64 speakers, 31 of which were diagnosed with ALS. Each speaker was asked to produce sustained phonation of vowels /a/ and /i/ at a comfortable pitch and loudness as constant and long as possible. It can be seen that voice database is almost balanced and contains 48\% of pathological voices and 52\% of healthy voices.

The age of the 17 male patients ranges from 40 to 69 (mean 61.1$\;\pm\;$7.7) and the age of the 14 female patients ranges from 39 to 70 (mean 57.3$\;\pm\;$7.8). For the case of healthy controls (HC), the age of the 13 men ranges from 34 to 80 (mean 50.2$\;\pm\;$13.8) and the age of the 20 females ranges from 37 to 68 (mean 56.1$\;\pm\;$9.7). The samples were recorded at 44.1 kHz using different smartphones with a regular headsets and stored as 16 bit uncompressed PCM files. Average duration of the records in the HC group was 3.7$\;\pm\;$1.5 s, and in ALS group 4.1$\;\pm\;$2.0 s. The detailed information about ALS patients is presented in table~\ref{tbl:participant}. All the patients were judged by the neurologist (the second author) to have presence of bulbar motor changes in speech (last column of the table~\ref{tbl:participant}).

\begin{table}[th]\centering 
\caption{ALS participants clinical records} 	
\footnotesize
\begin{tabular}{lccccc}
\hline 
\begin{minipage}[l]{12mm} Subject code \end{minipage} & 
Sex & 
Age &  
\begin{minipage}[c]{20mm} \centering\vspace{1mm} Time from ALS onset (months)\vspace{1mm} \end{minipage}  & 
\begin{minipage}[c]{20mm} \centering Bulbar/ spinal onset \end{minipage} & 
\begin{minipage}{20mm} \centering Presence of the bulbar signs \end{minipage} \\
\hline
008 & M & 67& 28 & bulbar & yes\\ 
020 & F  & 57 & 35 & spinal & no\\ 
021 & F  & 55 & 15 & spinal & yes\\ 
022 & F  & 70 & 11 & bulbar & yes\\ 
024 & M & 66 & 16& spinal & no \\
025 & M & 51 & 7  & spinal & no \\
027 & M & 57 & 18& bulbar& yes\\
028 & M & 58 & 5  & spinal& yes\\
031 & M & 67 & 6  & spinal& yes\\
032 & M & 61 & 19& spinal& yes\\
039 & M & 67 & 12& bulbar& yes\\
042 & M & 67 & 22& spinal& yes\\
046 & F & 50  & 12& spinal& yes\\
048 & F & 63  & 22& bulbar& yes\\
052 & F & 62  & 36& spinal& no \\
055 & M & 61& 11& spinal& yes\\
058 & M & 58&  9&  bulbar& yes\\
062 & M & 57& 23& bulbar& yes\\
064 & M & 57& 58& spinal& yes\\
068 & M & 40& 11& bulbar & yes\\
072 & F & 64 & 10& spinal& yes\\
076 & M & 68& 12& bulbar & yes\\
078 & F & 64 & 12& bulbar & yes\\
080 & F & 63 & 20& bulbar & yes\\
084 & F & 55 & 33& bulbar & yes\\
092 & F & 39 & 57& spinal & no \\
094 & F & 55 & 14& spinal& no \\
096 & F & 52 & 14& spinal& yes\\
098 & M & 68 & 37&spinal& yes\\
100 & M & 68 & 16& bulbar & yes\\
102 & F & 53 & 25& spinal& no\\
\hline
\end{tabular}
\label{tbl:participant}
\end{table}

\subsection{Aggregation of feature set and its statistical survey}
\begin{figure}[htb]
    \centering
     \includegraphics[width=0.64\textwidth]{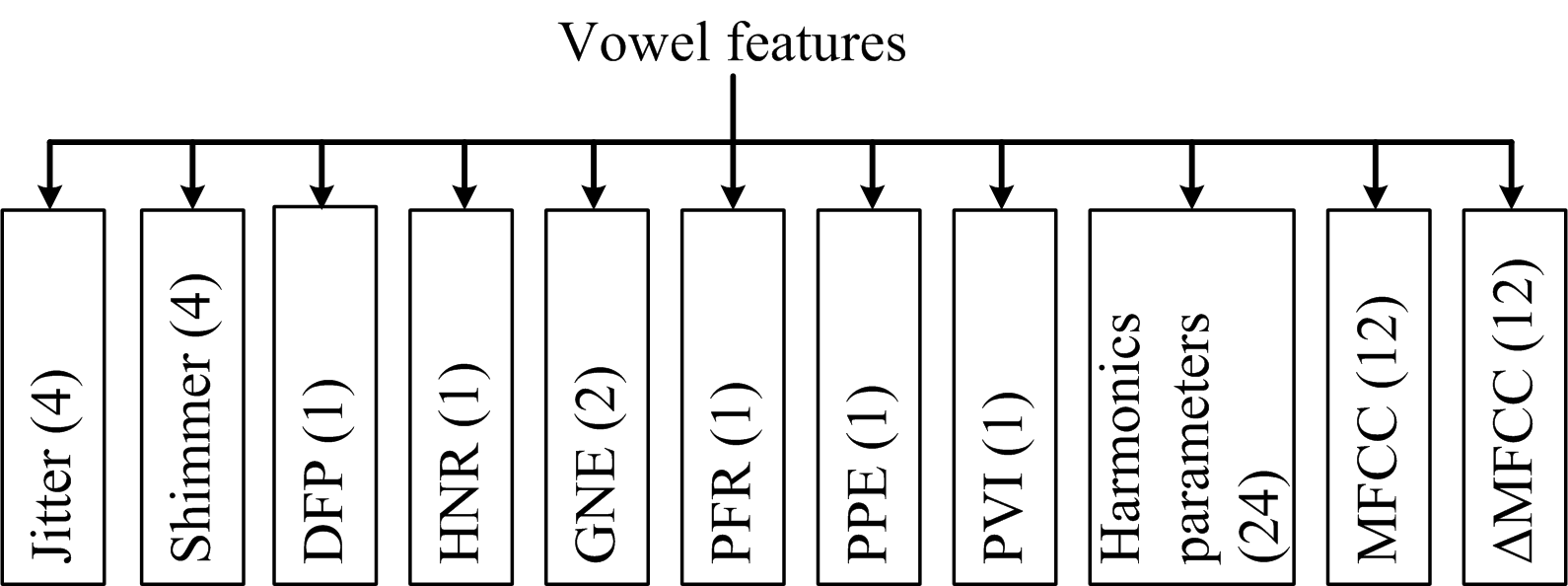} 
\caption{Features extracted from SVP test of one vowel}
\label{fig:feature_vector}    
\end{figure}

For each vowel used in SVP test 64 features are extracted (see figure~\ref{fig:feature_vector}). These features include the
following groups (the number of parameters in each group is indicated in parentheses): jitter (4), shimmer (5), DPF(1),  HNR (1), GNE (mean and SD), PFR(1), PPE(1), PVI (1),  $\mathrm{H}p^\mu$ (8), $\mathrm{H}p^\sigma$ (8), $\mathrm{RelH}p$ (8),  MFCC (12), $\Delta$~MFCC (12).
We also used three additional parameters  $d_1(E_a,E_i)$, $F2_{conv}$ and $F2_i$ (extra feature for vowel /i/). Thus the total number of features used in this study was 131 (64 for vowel /a/, 64+1 for /i/ and 2 joint parameters). 
In most cases we have used lower subscript to indicate the vowel for which feature was calculated. For example, $\mathrm{H2}^{\sigma}_i $ is SD of 2nd harmonic of vowel /i/ phonation.
 
In order to get initial understanding of the statistical properties of the features, we computed the Pearson correlation coefficient $r(\mathbf{x},\mathbf{y})$, where the vector $\mathbf{x}$ contains the values of a single feature for all phonations, and $\mathbf{y}$ is the associate labels (``$0$''  for healthy subject, ``$1$'' -- for ALS patient).

\subsection{Feature selection}
\label{sec:FS}
It is known that reducing the number of features often improves the model's predictive power. Also the reduced feature subset give better insight into the problem via analysis of the most predictive features~\cite{Flach-2012}.

In this study we used four efficient feature selection (FS) approaches: 1) maximization of quality of variation (QoV)~\cite{Liu-11}, 2) Relief~\cite{Kira-92} 3) least absolute shrinkage and selection operator (LASSO)~\cite{Tibshirani-94}, 4) RelieFF~\cite{Kononenko-1997}. Maximization of QoV is a noise-resistant method for feature selection based on order statistics. The basic notion of this method is {\it class impurity} -- characteristic that is calculated for each feature based on its order statistics. The {\it quality of variation} of a feature is inverse of the average impurity of all the classes along the feature. This method allows one to rank all features according to the QoV criterion. It has been show that QoV method performs well when the available training data is small or not much bigger compared to the dimensionality of feature vector~\cite{Liu-11}. 
LASSO is a linear regression based technique that minimizes the residual sum of squares subject to the absolute value of the coefficient being less than a constant. This leads to some coefficients that are shrunk to zero, which in essence means that feature associated with those coefficients are eliminated. In order to rank the features using LASSO we repeat its computation with different values of regularization parameter $\lambda$ in order to track the order in which features are eliminated. The first eliminated feature is considered as least informative while the last as the most relevant.
The key idea of Relief is to estimate features according to how well their values distinguish among the instances that are near to each other. Original Relief algorithm estimates relevance of feature for a given instance by analysis closest neighbors: one from the same class (nearest hit) and one from the opposite class (nearest miss). Advanced version RelieFF extends this idea to k nearest neighbors. 
Overall, all four feature selection algorithms have shown promising results in machine learning application. 

\subsection{Classification}
For a binary classification between normal and pathological classes, linear discriminant analysis (LDA) with Fisher criterion was used~\cite{hastie-01}. The basic idea of LDA consists in searching for such a direction $\mathbf{w}$ in the feature space, that the projection of all training vectors onto it minimizes the within-class variation and maximizes the between-class variation:
\begin{equation}
\mathbf{w} = \arg\max_{\mathbf{w}}\frac{\mathbf{w S_{\mathit{B}} w}^T}{\mathbf{w S_{\mathit{W}} w}^T},
\label{eq:hyper_w}
\end{equation}
where $\mathbf{S_{\mathit{B}}}$ -- between class scatter matrix and $\mathbf{S_{\mathit{W}}}$ -- within class scatter matrix. In turn these matrices are calculated as follows
\begin{equation}
\mathbf{S_{\mathit{B}}} = (\mu_1 - \mu_2)(\mu_1 - \mu_2)^T,
\label{eq:S_b}
\end{equation}
\begin{equation}
\mathbf{S_{\mathit{W}}} =\sum_{j=1}^{2}\sum_{\mathbf{x}}(\mathbf{x} - \mu_j)(\mathbf{x} - \mu_j)^T,
\label{eq:S_w}
\end{equation}
where $\mathbf{x}$ -- feature vectors from training set,  $\mu_1$ -- mean value of feature vector for healthy people and $\mu_2$ -- mean value of feature vector for people with ALS. The solution of (\ref{eq:hyper_w}) can be found via the generalized eigenvalue problem 
\begin{equation}
\mathbf{S_{\mathit{B}}} \mathbf{w} = \lambda_m \mathbf{S_{\mathit{W}}} \mathbf{w},
\label{eq:eig_LDA}
\end{equation}
where the eigenvector associated with maximum eigenvalue $\lambda_m$ gives the projection basis. Classification function of LDA is formulated as follows
\begin{equation}
f(\mathbf{x}) = \mathrm{sign}\bigl(\bigl<\mathbf{w},\mathbf{x}\bigr>+b\bigr),
\label{eq:LDA_fun}
\end{equation}
where $b$ is a bias. In the experiments, the value of $b$ was chosen in a such way that the number of correctly detected positive and negative instance in the training set was equal.  More detailed description of LDA can be found in~\cite{hastie-01}.

\subsection{Classifier Validation}
\label{sec:valid}
The goal of validation is to estimate of the generalization performance of the classification based on the selected set of features, when presented new (previously unseen) data. Most studies use cross-validation to achieve this goal~\cite{Tsanas-2012,Orozco-15,Cordeiro-2018}.

In this work we used $k$-fold stratified cross validation (CV) method~\cite{Kohavi-95}, with $k$ equal to 8. According to this method at the beginning of the CV process dataset randomly permuted and then splits into eight equal subsets (folds) ($s_1$--$s_8$), the folds are stratified so that they contain approximately the same proportions of labels as original dataset. At first iteration classifier is trained using subsets $s_1$--$s_7$, while testing is conducted using $s_8$ subset. Then training is repeated using $s_2$--$s_8$ subsets, and classifier tested using $s_1$ subset, and so on. After 8 iteration whole dataset is labelled using eight classifiers.
This process was repeated a total of 40 times. The classification performance is evaluated in terms of the mean and standard deviation of the accuracy on the test set across all folds. 

Accuracy, sensitivity, and specificity were used in this study to measure the classification performance. Accuracy is the overall probability of correctly classified instance over the total number of instances. Sensitivity is the probability of correctly classified ALS patients given all ALS samples and specificity is probability of classified HC given all HC samples. Accuracy, sensitivity, and specificity are calculated as follows:
\begin{eqnarray*}
Acc &=&\displaystyle \frac{TP+TN}{TP+FP+FN+TN} \\
Sens &=& \displaystyle \frac{TP}{TP+FN} \\
Spec &=& \displaystyle \frac{TN}{TN+FP}
\end{eqnarray*}
where $TP$, $TN$, $FP$, $FN$ -- the number of true positive, true negative, false positive and false negative results of classification, respectively. In this case, positive means a prediction that the voice sample is produced by a speaker with ALS.

\section{Results}
\label{sec:results}
\subsection{Preliminary statistical survey}

Table~\ref{tbl:stat} presents the several features most strongly associated with the labels in dataset, sorted by the absolute value of the correlation coefficient~\cite{hastie-01}. We used label ``$0$'' for healthy controls and ``$1$'' for people with ALS. Thus, positive correlation coefficient suggest that the feature takes, in general, larger value for ALS voices. All of the listed  features exhibit statistically significant correlation ($p<0.05$). 

\begin{table}[th]\centering 
\caption{Statistical analysis of the acoustic features} 	
\begin{tabular}{lcc}
\hline \hline
Feature   & 
\begin{minipage}[c]{25mm} \centering Correlation coefficient \end{minipage} & \begin{minipage}{25mm} \centering\vspace{1mm} Difference between groups \vspace{1mm}\end{minipage} \\
\hline
$d_1$                                      & $-$0.456    & $p<0.0002$ \\
$\mathrm{MFCC}_{i}(2)$  & $-$0.446    & $p<0.0003$ \\
$\mathrm{PVI}_{a}$            &       0.422    & $p<0.0006$ \\ 
$\mathrm{PPE}_{a}$           &       0.418 	  & $p<0.0006$ \\
$F2_{conv}$                          & $-$0.390    & $p<0.002$  \\ 
$\mathrm{RelH7}_a$           & $-$0.381    & $p<0.002$ \\ 
$\mathrm{MFCC}_{i}(6)$  &       0.371    & $p<0.003$  \\ 
$J_{ppq55}^{(a)}$               &       0.361    & $p<0.004$  \\ 
$\mathrm{PVI}_{i}$            &       0.351    & $p<0.005$ \\  
$\mathrm{RelH1}_i$  	     & $-$0.347    & $p<0.005$ \\  
$\mathrm{PFR}_{a}$          &       0.346     & $p<0.006$  \\ 
$\mathrm{H8}_a^{\mu}$   &  $-$0.335    & $p<0.007$ \\ 
$\mathrm{GNE}_a^{\mu}$ & $-$0.324    & $p<0.01$ \\ 
$\Delta\mathrm{MFCC}_{i}(6)$  & 0.321& $p<0.01$ \\ 
$S^{(i)}_{apq11}$               &       0.311    & $p<0.02$ \\ 
$F2_i$    						          & $-$0.302    & $p<0.02$ \\ 
$\mathrm{RelH1}_a$  	     & $-$0.285    & $p<0.03$ \\ 
$\mathrm{GNE}_i^{\sigma}$&    0.282    & $p<0.03$ \\ 
$\mathrm{H4}_a^{\sigma}$&       0.282    & $p<0.03$ \\ 
$\mathrm{MFCC}_{i}(8)$  &       0.273    & $p<0.03$\\ 
$\mathrm{MFCC}_{a}(11)$&      0.250    & $p<0.05$\\ 
\hline
\end{tabular}
\label{tbl:stat}
\end{table}

According to the table~\ref{tbl:stat} the most relevant features are $d_1$ and $\mathrm{MFCC}_{i}(2)$. The distance between spectral envelopes of the vowels /a/ and /i/ ($d_1$) has the strongest correlation with the labels in the dataset. The negative sign of its correlation coefficient means that the smaller distance $d_1$, the more likely that voice belongs to the category of ALS patients. $\mathrm{MFCC}_{i}(2)$ has almost the same strong correlation as spectral distance $d_1$. It is well known that, low-order MFCC describes the spectral envelope of the sound, therefore it can be concluded that patients with ALS usually have significant changes in spectral envelope of the vowel /i/.

Parameters $\mathrm{PVI}_{a}$   and $\mathrm{PPE}_{a}$  (third and fourth rows in the table~\ref{tbl:stat}) also have high correlation with the labels in the dataset. This fact indicates that, as a result of neuromotor disorders in patients with ALS, oscillations uncharacteristic for healthy people appear in the F0 contour, which lead to increasing of $\mathrm{PVI}_{a}$ and $\mathrm{PPE}_{a}$.
It is interesting that along with $\mathrm{PVI}_{a}$ parameter $\mathrm{PVI}_{i}$ (ninth row) is also presented in table~2 and has high correlation coefficient, while $\mathrm{PPE}_{i}$ does not show a statistically significant correlation ($p>0.14$). This may indicate that $\mathrm{PVI}$ less depends on type of analyzing vowel than $\mathrm{PPE}$ and better reflects changes associated with a decrease in the control over fundamental frequency in patients with ALS.

Five features ($\mathrm{RelH7}_a$, $\mathrm{RelH1}_i$, $\mathrm{H8}_a^{\mu}$,  $\mathrm{RelH1}_a$ and $\mathrm{H4}_a^{\sigma}$), among twenty-one of those listed in Table 2, relate to parameters that describe harmonic structure of the vowel. This suggests that these parameters could be useful for accurate voice classification.

\begin{figure}[tbh]
    \centering
    \begin{minipage}{0.49\textwidth}
        \centering
        \includegraphics[width=0.99\textwidth]{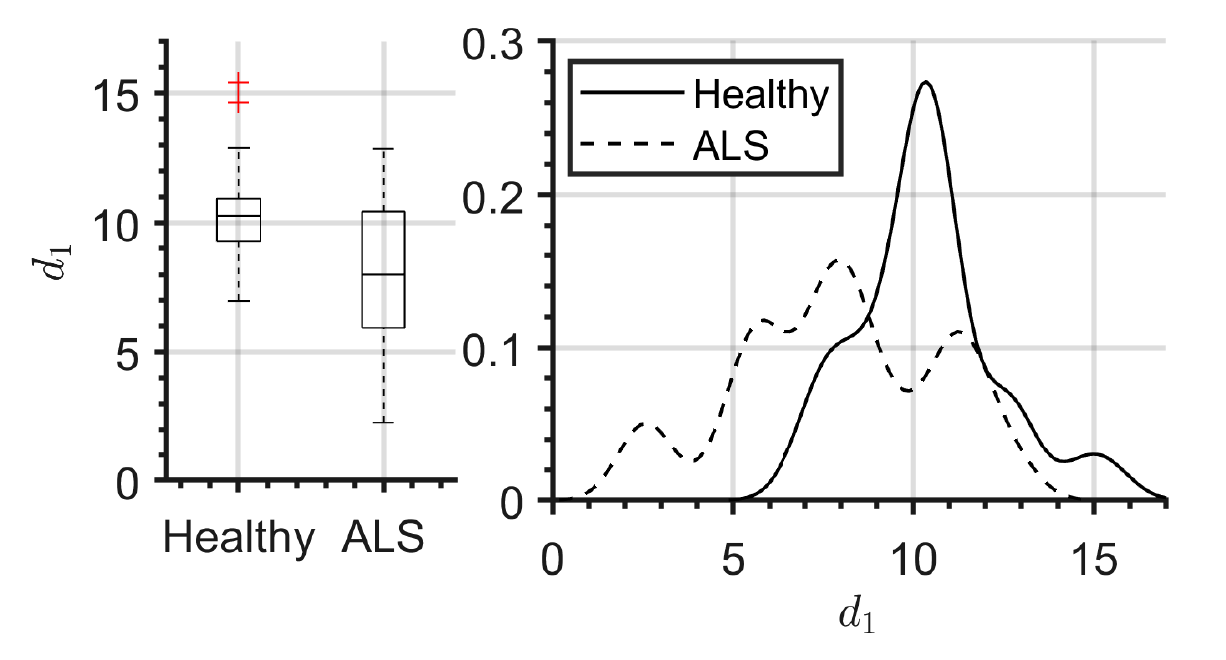} 
       \small (a)
    \end{minipage}\hfill
    \begin{minipage}{0.49\textwidth}
        \centering
        \includegraphics[width=0.99\textwidth]{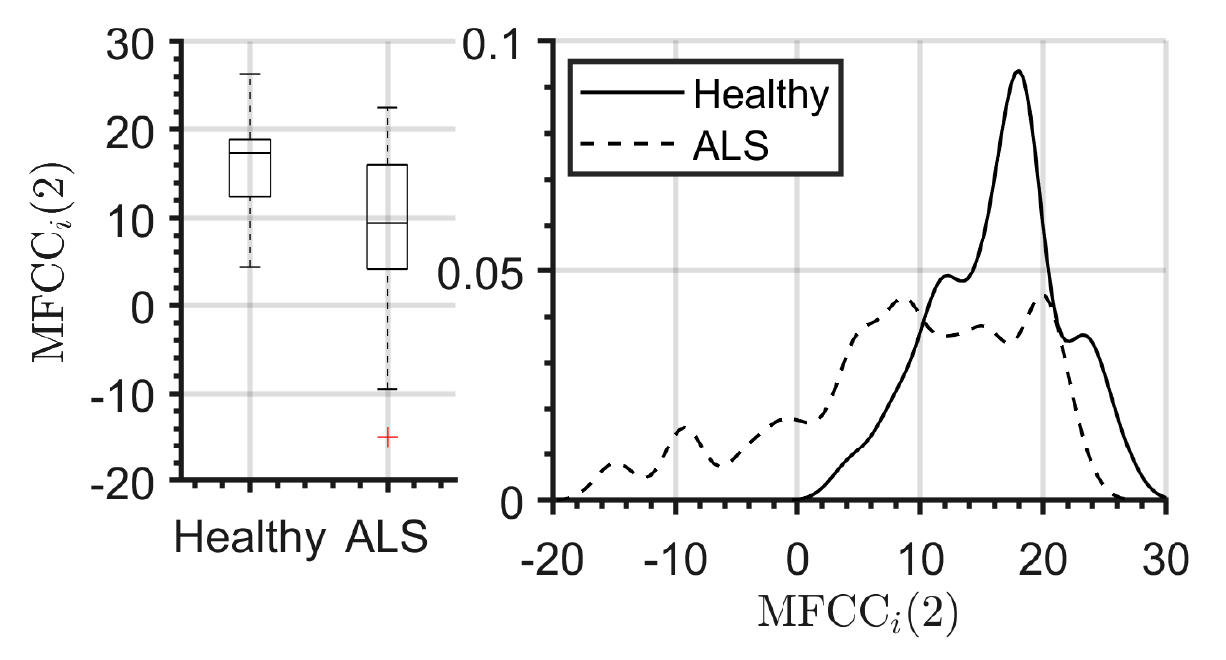} 
       \small (b)
    \end{minipage}
\caption{Box plot and probability densities of (a) $d_1(E_a,E_i)$; (b)$\mathrm{MFCC}_i(2)$ }
\label{fig:d1_mfcc_i2}    
\end{figure}

We also estimated distribution of several features listed in table~\ref{tbl:stat} using Gaussian kernel density method to characterize their statistical properties. Figure~\ref{fig:d1_mfcc_i2},a shows the distribution of distance between spectral envelopes of the vowels /a/ and /i/ (first row in table~\ref{tbl:stat}). As expected, on average this feature has lower value for ALS patients than for healthy subjects.

\begin{figure}[th]
    \centering
    \begin{minipage}{0.49\textwidth}
        \centering
        \includegraphics[width=0.99\textwidth]{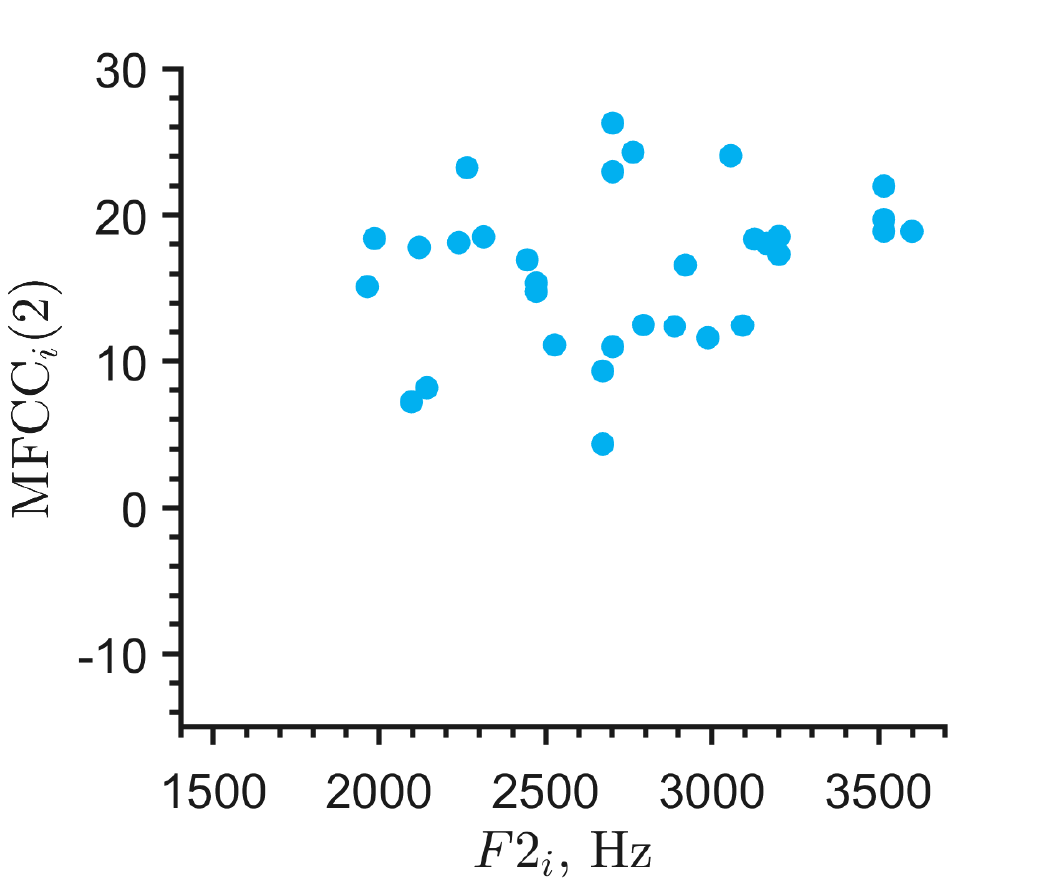} 
       \small (a)
    \end{minipage}\hfill
    \begin{minipage}{0.49\textwidth}
        \centering
        \includegraphics[width=0.99\textwidth]{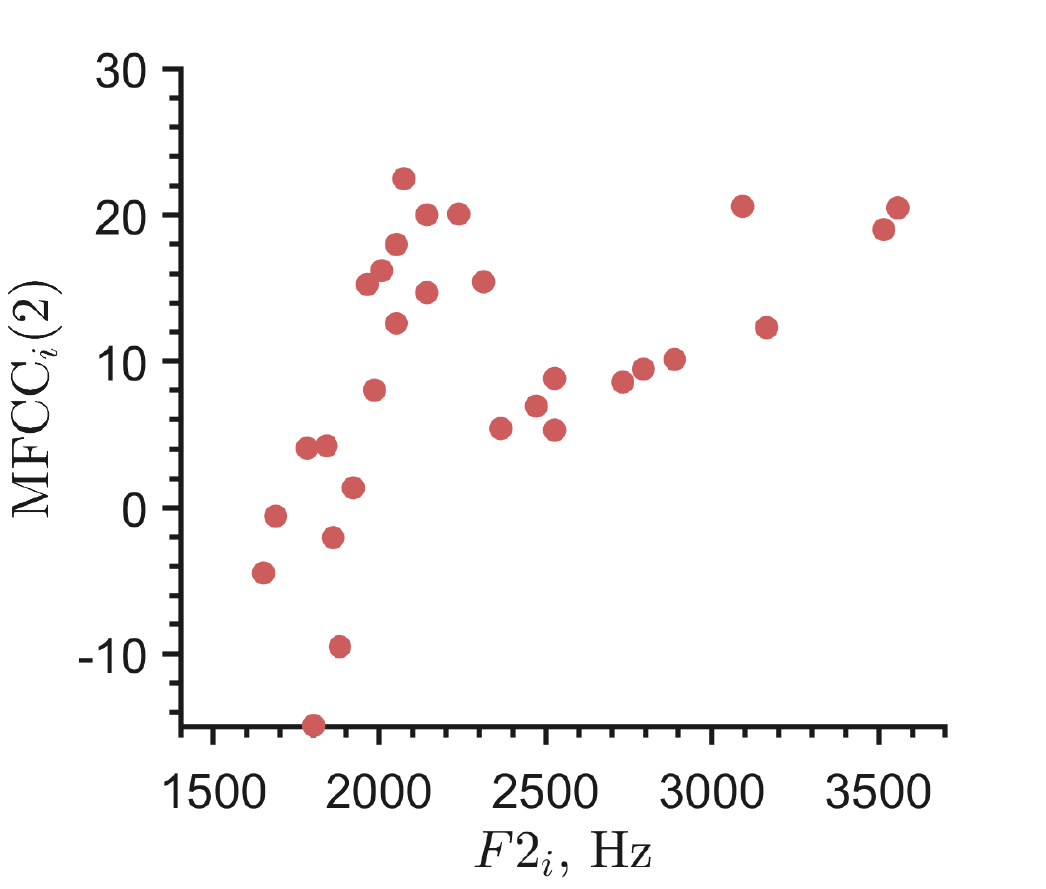} 
       \small (b)
    \end{minipage}
\caption{Scatter plots of pairs of $F2_i$ and $\mathrm{MFCC}_i(2)$ showing low correlation for healthy voices (a) and high correlation for ALS voices (b)}
\label{fig:scatter}    
\end{figure}

The distribution of 2nd MFCC of vowel /i/ that has a strong correlation with labels in the dataset is given in figure~\ref{fig:d1_mfcc_i2},b.  As stated above the differences in $\mathrm{MFCC}_{i}(2)$ indicate  changes in the spectral envelope of the vowel /i/ in patients with ALS. Among the others, this can be seen from the changes of the second formant frequency of the vowel /i/. From table 2 we see that a lower value of $F2_i$ is typical for patients with ALS. This observation is consistent with previous findings in this area~\cite{Lee-2017,Lee-2019}. Let us consider scatter plot of the pairs of $F2_i$ and $\mathrm{MFCC}_{i}(2)$ for healthy and pathological voices (see figure~\ref{fig:scatter}). It can be seen that for healthy voices $F2_i$ and $\mathrm{MFCC}_{i}(2)$ are weakly correlated (i.e. they not set out along slanting line). In contrast, for the voices of patients with ALS, it can be seen that $F2_i$ and $\mathrm{MFCC}_{i}(2)$ are strongly correlated (points are grouped along slanting line). Thus high relevance of the $\mathrm{MFCC}_{i}(2)$ is likely caused by the fact that it reflects the changes in second formant of vowel /i/ in patients with ALS.

 
\begin{figure}[tbh]
    \centering
    \begin{minipage}{0.49\textwidth}
        \centering
        \includegraphics[width=0.99\textwidth]{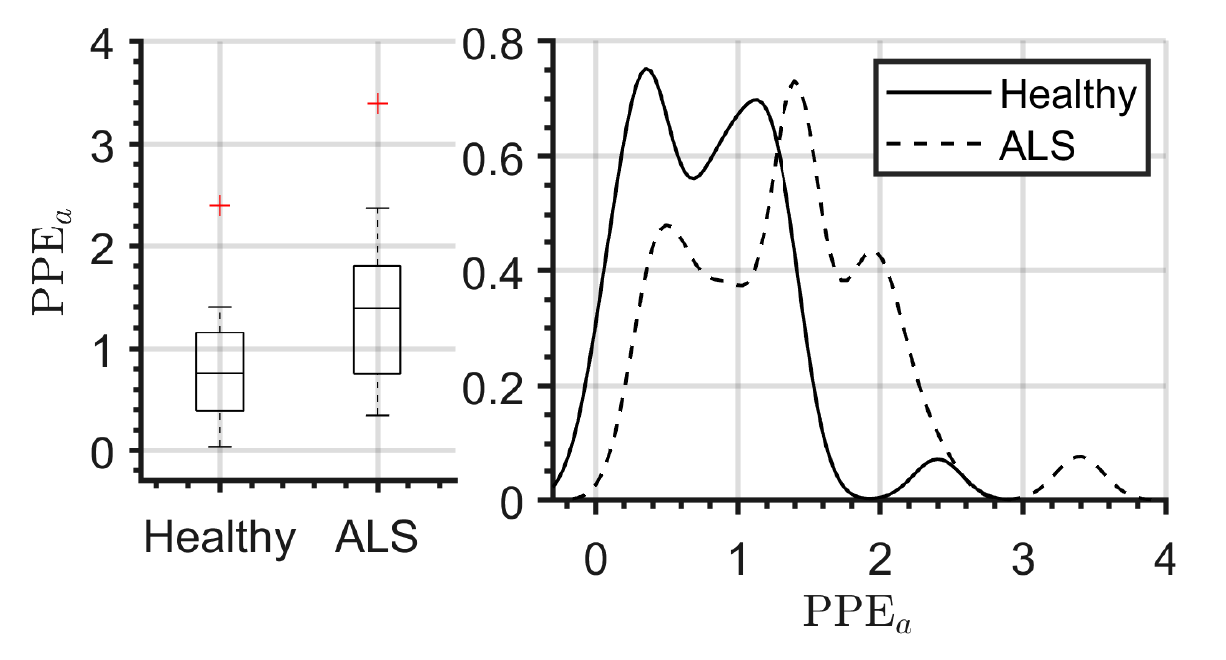} 
       \small (a)
    \end{minipage}\hfill
    \begin{minipage}{0.49\textwidth}
        \centering
        \includegraphics[width=0.99\textwidth]{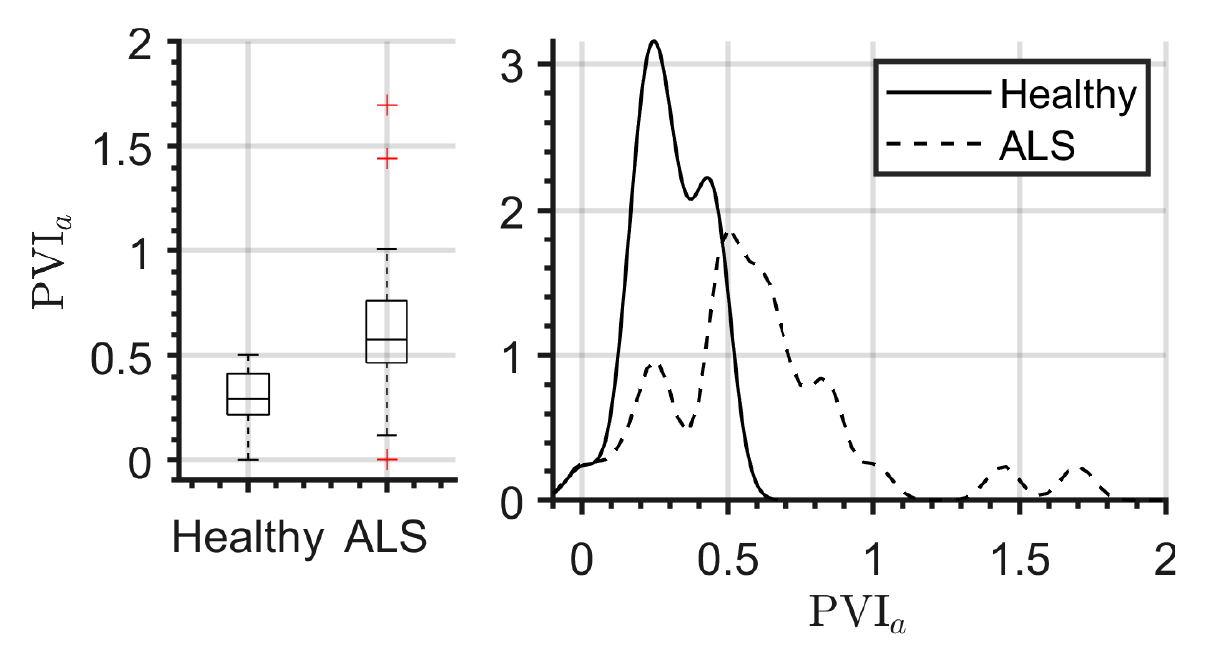} 
       \small (b)
    \end{minipage}
    \\
     \begin{minipage}{0.5\textwidth}
          \centering
          \includegraphics[width=1.0\textwidth]{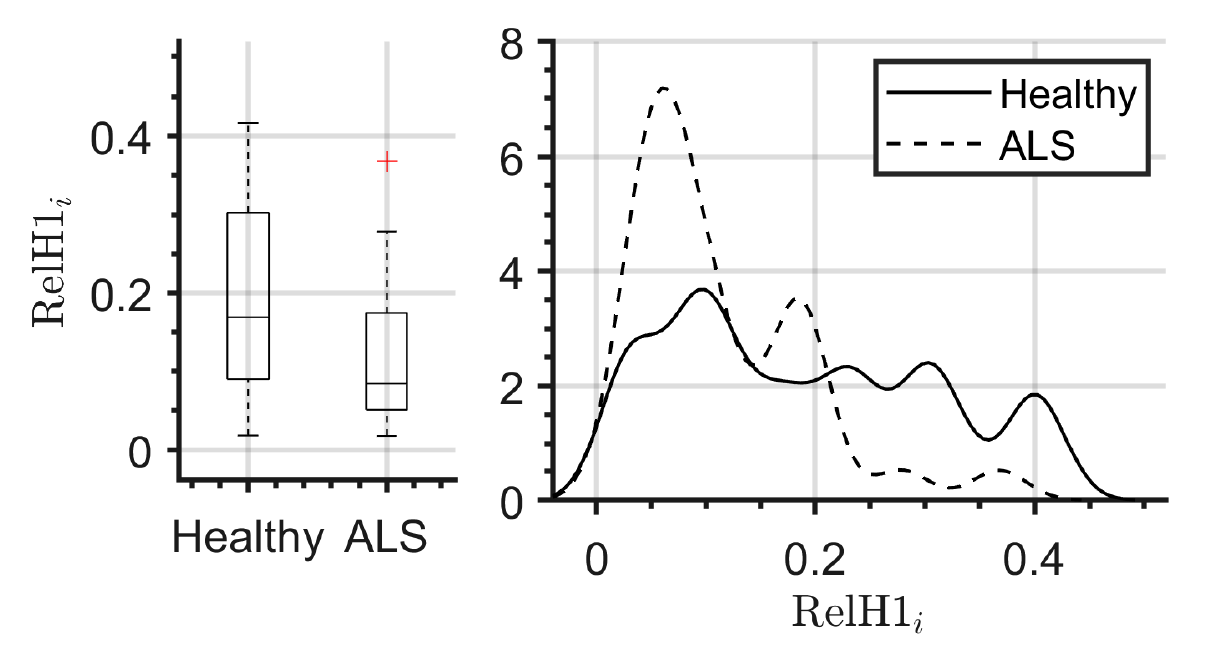} 
          \small (c)
        \end{minipage}\hfill
        \begin{minipage}{0.5\textwidth}
          \centering
          \includegraphics[width=1\textwidth]{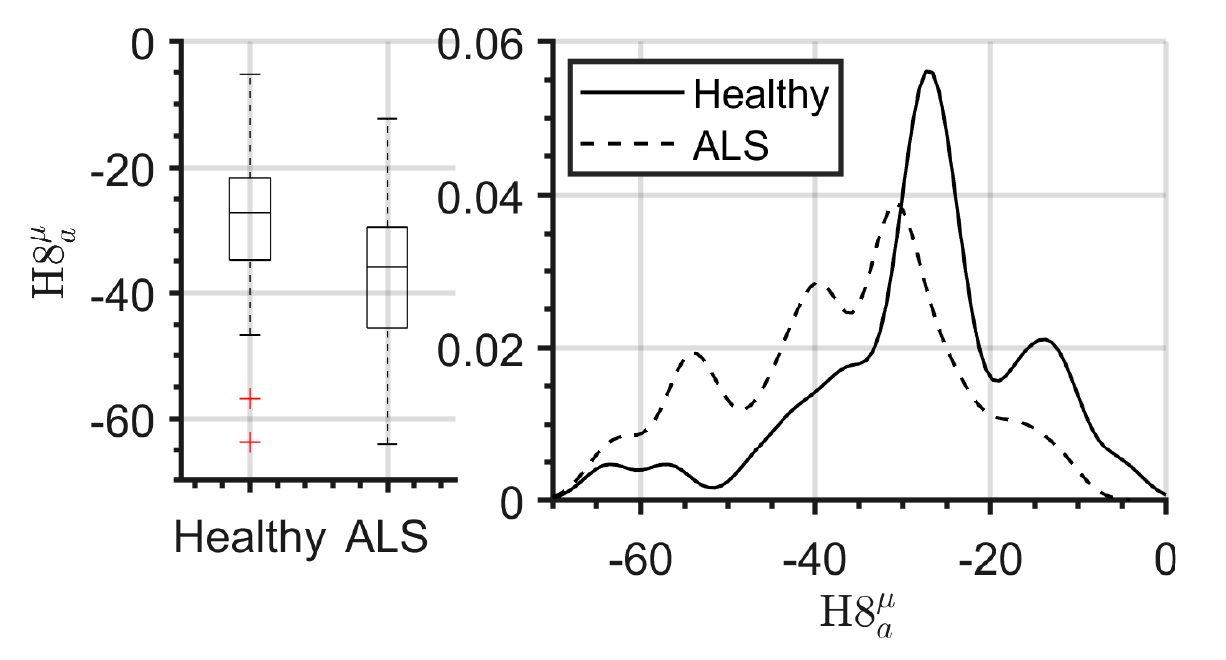} 
          \small (d)
        \end{minipage}
        \begin{minipage}{0.5\textwidth}
        		\centering
        		\includegraphics[width=1.0\textwidth]{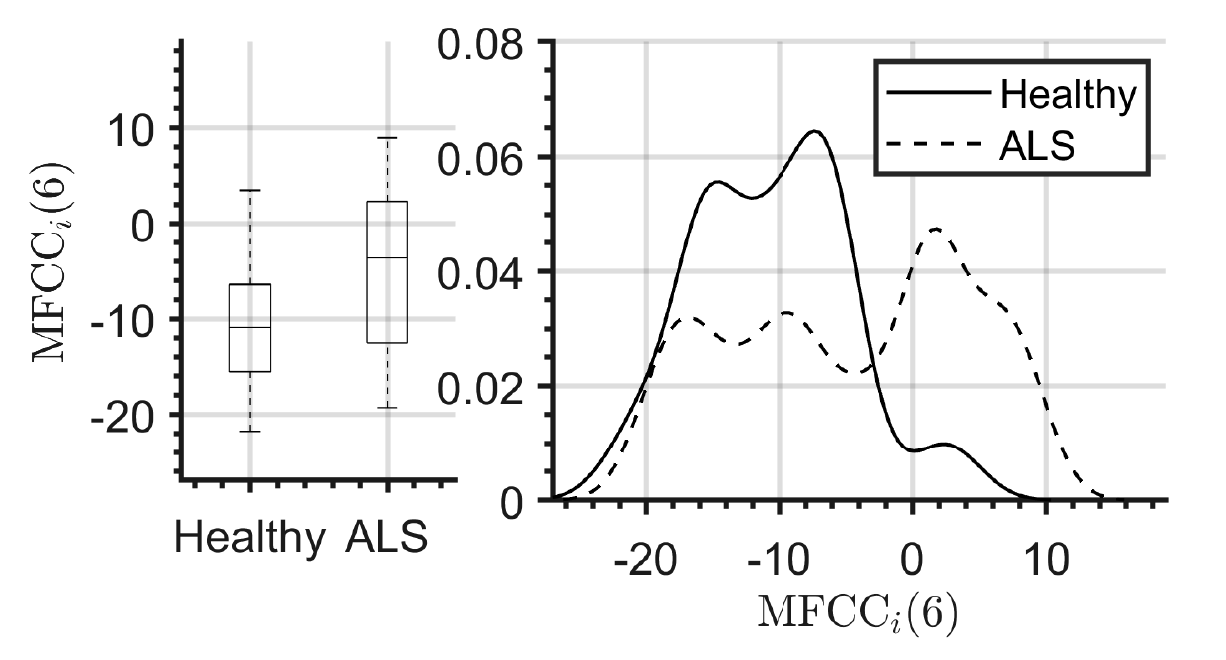} 
        		\small (e)
        	\end{minipage}\hfill
        	\begin{minipage}{0.5\textwidth}
        		\centering
        		\includegraphics[width=1\textwidth]{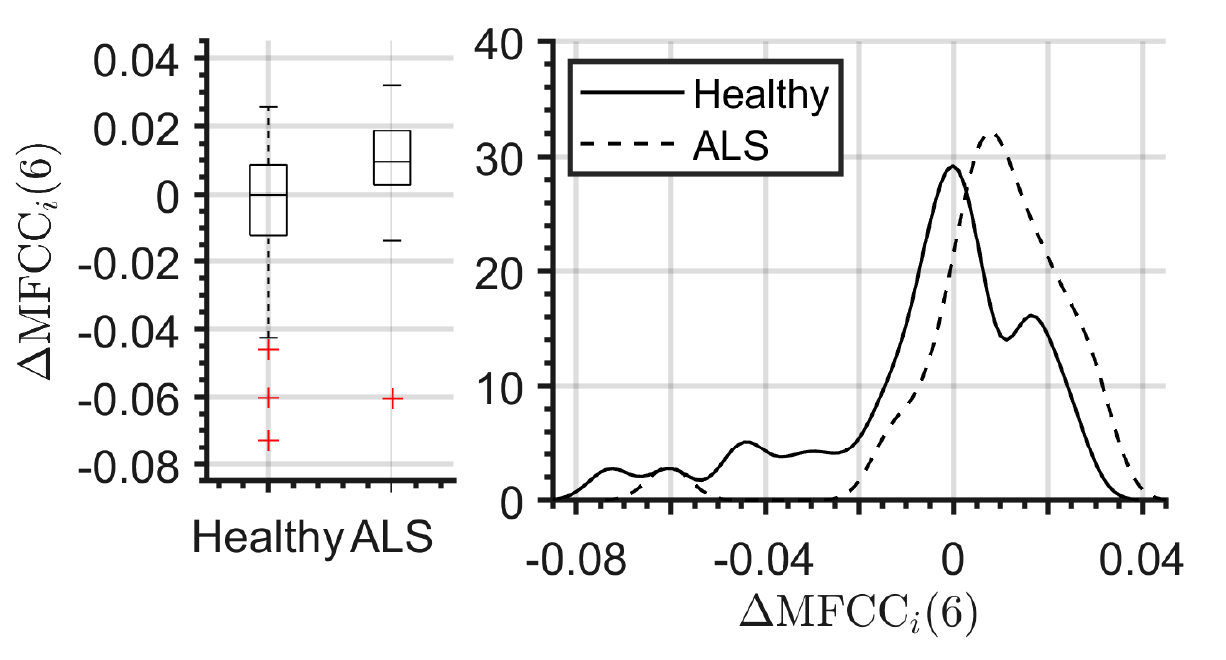} 
        		\small (f)
        	\end{minipage}
\caption{Box plot and probability densities of : (a) $\mathrm{PPE}_a$; (b) $\mathrm{PVI}_a$;  (c) $\mathrm{RelH1}_i$; (d) $\mathrm{H8}_a^{\mu}$; (e) $\mathrm{MFCC}_i(6)$; (f) $\Delta\mathrm{MFCC}_i(6)$}
\label{fig:box_plots}    
\end{figure}
Figure~\ref{fig:box_plots},a-b illustrate distributions of $\mathrm{PPE}_a$ and $\mathrm{PVI}_a$ features. Both of them  characterize the excess of variability in a pitch contour and have high correlation coefficients (3-rd and 4-th rows of table~\ref{tbl:stat}).
Comparing boxplots of the  $\mathrm{PPE}_a$ and $\mathrm{PVI}_a$ parameters, we can see that the first quartile of $\mathrm{PPE}_a$ for pathological voices is located at the level of the median of the $\mathrm{PPE}_a$ for healthy voices. In turn, the first quartile of $\mathrm{PVI}_a$ for pathological voices exceeds the third quartile of $\mathrm{PVI}_a$ for healthy voices. This indicates that $\mathrm{PVI}_a$ has stronger discriminatory power than $\mathrm{PPE}_a$.

Examples of features that characterized harmonic structure of voice are given in figure~\ref{fig:box_plots},c-d. Both features are strongly associated with the labels in the dataset. As expected, distributions figure~\ref{fig:box_plots},c indicate that $\mathrm{RelH1}_i$ has tending to have higher value for healthy voices. Figure~\ref{fig:box_plots},d shows that 8-th harmonic of vowel /a/ has lower mean value in ALS group.


Figure~~\ref{fig:box_plots},e-f illustrates the distributions of the MFCC and delta MFCC that have high correlation with labels in the dataset (rows 7 and 14 in table~\ref{tbl:stat}).  Boxplot in figure~\ref{fig:box_plots},f shows that $\Delta\mathrm{MFCC}_i(6)$ have almost symmetrical distribution with median greater than zero for ALS voices, while for healthy voices this parameter have asymmetrical distribution with near zero median.

%

The presented findings give tentative confidence that we can expect good results for the classification problem of this study. 

\subsection{Classification results and discussion}
\begin{figure}[tbh]
  \centering
  \includegraphics[width=130mm]{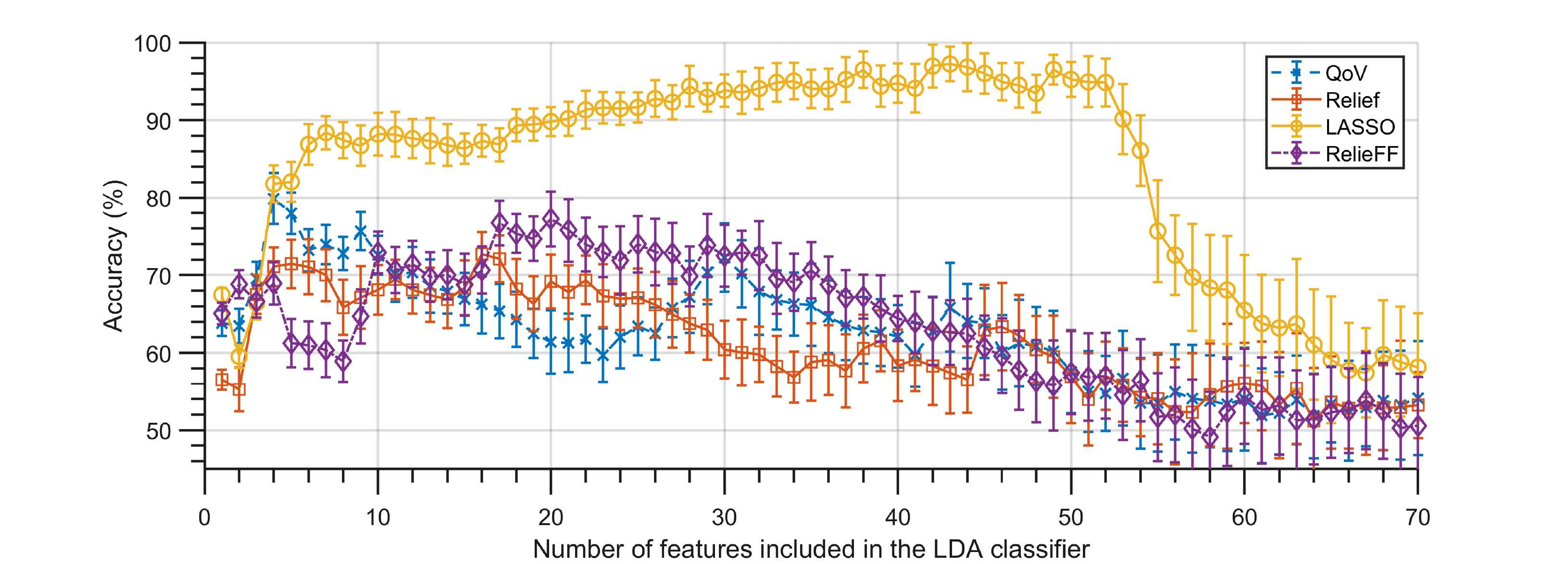}
  \caption{Classification accuracy with confidence interval (one standard deviation around the quoted mean accuracy). The results obtained using different feature selection algorithm. For RelieFF algorithm adjustable parameter $k=11$ was used.
  }
  \label{fig:perf_ai}
\end{figure}

In our experiments we computed the accuracy (see section~\ref{sec:valid}) of LDA classifier using different number of features selected by the four FS algorithms described in section~\ref{sec:FS}. Figure~\ref{fig:perf_ai} shows the obtained results. 

The analysis of figure~\ref{fig:perf_ai} shows that performance of all FS algorithm is quite similar while the number of features $N$ is less than 6. However, for $N>6$ LASSO demonstrates significantly better performance in comparison with other approaches. Possible explanation of this fact is that mathematical principles of LASSO regression are in accordance with the discriminant function (\ref{eq:LDA_fun}) of the LDA classifier.

The optimal size of the feature vector is equal to 43 and it was achieved using the LASSO approach.
The accuracy obtained in this case is 97\%. This result considerably outperforms the others. For example, the best accuracy of LDA classifier with QoV FS algorithm is 79\% and was achieved using 4 features. The best results obtained with RelieFF and Relief algorithms are even smaller -- 76\% and 72\%, accordingly.

It is always desirable to have a classifier with a low number of features, therefore we applied  backward-step selection procedure~\cite{Flach-2012} to reduce the number of features picked out by FS algorithms. The backward-step selection starts with LDA model that used best feature subset found by FS algorithm, and sequentially deletes the feature that has  low (or negative) impact on the fit. The result of the described  feature selection process is summarized in~table~\ref{tbl:perf}. 

\begin{table}[tbh]\centering 
\caption{Classifiers accuracy obtained using different feature selection (FS) algorithms. The resulting number of features is given in parentheses.} 	
\renewcommand{\arraystretch}{1.2} 
\begin{tabular}{|l|c|c|}
\hline \hline
FS  algorithm  &  \begin{minipage}[c]{30mm}\vspace{1mm}\centering Accuracy with \\ initial subset \end{minipage} & \begin{minipage}[c]{30mm}\vspace{1mm}\centering Accuracy after backward- \\ stepwise selection \end{minipage} \\
\hline
QoV   	& \begin{minipage}{30mm}\centering \vspace{2mm} $79.5\pm3.5$\%  \\  $(4)$ \vspace{2mm} \end{minipage}   &  \begin{minipage}{30mm}\centering \vspace{2mm} $79.5\pm3.5$\%  \\  $(4)$ \vspace{2mm} \end{minipage}   \\
Relief  &\begin{minipage}{30mm}\centering \vspace{2mm} $72.5\pm3.4$\%  \\  $(16)$ \vspace{2mm} \end{minipage}   &  \begin{minipage}{30mm}\centering \vspace{2mm} $80.3\pm2.3$\%  \\  $(5)$ \vspace{2mm} \end{minipage}   \\ 
LASSO   & \begin{minipage}{30mm}\centering \vspace{2mm}\bf  97.0$\pm$2.4\%  \\  $(43)$ \vspace{2mm} \end{minipage}   &  \begin{minipage}{30mm}\centering \vspace{2mm}\bf 99.7$\pm$0.6\%  \\  $(32)$ \vspace{2mm} \end{minipage}   \\ 
RelieFF & \begin{minipage}{30mm}\centering \vspace{2mm} $75.9\pm4.2$\%  \\  $(20)$ \vspace{2mm} \end{minipage}   &  \begin{minipage}{30mm}\centering \vspace{2mm} $82.9\pm2.8$\%  \\  $(11)$ \vspace{2mm} \end{minipage}   \\
\hline
\end{tabular}
\label{tbl:perf}
\end{table}

Result shown in table~\ref{tbl:perf} demonstrate that the best accuracy for LDA classifier is obtained using feature selected by the LASSO  algorithm. Also it can be noted that  backward-stepwise selection (BSS) is effective in reducing the number of features and increasing the accuracy of classifier. The most noticeable result, in this regard, is increasing the accuracy of LDA model with feature subset selected using RelieFF algorithm by 7 \%, while reducing the number of features by 9. Nevertheless, feature subsets found by QoV, Relief and RelieFF algorithms with application of BSS procedure give the resulting accuracy considerably lower compared to feature subset selected using LASSO algorithms. 

Table~\ref{tbl:features} lists final subsets of features selected using FS algorithms (with application of BSS).  The obtained accuracy, sensitivity and specificity for each cases are also given in table~\ref{tbl:features}.

\begin{table}[tbh]\centering 
\caption{Selected feature subsets and classification accuracy 
} 	
\renewcommand{\arraystretch}{1.2} 
\begin{tabular}{|l|c|c|}
\hline \hline
\begin{minipage}[c]{16mm}\vspace{2mm} FS \\ algorithm \vspace{1mm}\end{minipage}   	& Features & Accuracy results (\%) \\
\hline
\hline
\begin{minipage}[c]{16mm} QoV ($N=4$)
\end{minipage}   	& 
\begin{minipage}[c]{63mm}
$\mathrm{PVI}_a$,
$\mathrm{PFR}_a$,
$d_1$,
$\mathrm{MFCC}_i(2)$
\end{minipage}
& \begin{minipage}{30mm}\centering \vspace{2mm}  $Acc=79.5\pm3.5$ \\ $Sens=75.3\pm3.4$ \\ $Spec=83.5\pm5.9$ \vspace{2mm} \end{minipage}\\

\begin{minipage}[c]{16mm} Relief ($N=5$) \end{minipage}  &
\begin{minipage}[c]{63mm}
$\mathrm{PVI}_i$, 
$d_1$,
$\mathrm{MFCC}_i(4)$, 
$\mathrm{MFCC}_i(9)$,
$\Delta\mathrm{MFCC}_a(3)$
\vspace{2mm}
\end{minipage}
& \begin{minipage}{30mm}\centering\vspace{2mm}  $Acc = 80.3\pm2.3$ \\ $Sens = 68.1\pm3.4$ \\ $Spec = 91.7\pm2.3$ \vspace{2mm} \end{minipage}  \\

\begin{minipage}[c]{16mm} LASSO ($N=32$) \end{minipage}  & 
\begin{minipage}[c]{63mm}
  $\mathrm{PVI}_i$,
  $\mathrm{H2}_a^{\mu}$,
  $\mathrm{H4}_a^{\mu}$,    
  $\mathrm{H3}_a^{\sigma}$,  
  $\mathrm{RelH1}_a$,
  $\mathrm{RelH3}_a$,
  $\mathrm{RelH4}_a$,
  $\mathrm{RelH6}_a$,  
  $\mathrm{RelH8}_a$,    
  $\mathrm{RelH1}_i$,
  $\mathrm{RelH3}_i$,
  $\mathrm{MFCC}_a(1)$,
  $\mathrm{MFCC}_a(4)$,
  $\mathrm{MFCC}_a(7)$,  
  $\mathrm{MFCC}_a(10)$,
  $\mathrm{MFCC}_a(11)$,  
  $\mathrm{MFCC}_a(12)$,    
  $\Delta\mathrm{MFCC}_a(5)$,
  $\Delta\mathrm{MFCC}_a(9)$,
  $\Delta\mathrm{MFCC}_a(11)$,  
  $\mathrm{MFCC}_i(2)$,  
  $\mathrm{MFCC}_i(4)$,
  $\mathrm{MFCC}_i(8)$,
  $\mathrm{MFCC}_i(9)$,   
  $\Delta\mathrm{MFCC}_i(1)$,
  $\Delta\mathrm{MFCC}_i(9)$, 
  $\Delta\mathrm{MFCC}_i(10)$,
  $\Delta\mathrm{MFCC}_i(12)$
  $\mathrm{GNE}_a^{\sigma}$,   
  $\mathrm{GNE}_i^{\mu}$,   
  $\mathrm{GNE}_i^{\sigma}$,   
  $\mathrm{DPF}_a$
  \vspace{2mm}
\end{minipage}
& \begin{minipage}{30mm}\centering\vspace{2mm}  $Acc = \mathbf{99.7\pm0.6}$ \\ $Sens = \mathbf{99.3\pm1.4}$ \\ $Spec = \mathbf{99.9\pm0.5}$ \vspace{2mm} \end{minipage}  \\
\begin{minipage}[c]{16mm} RelieFF ($N=11$) \end{minipage} & 
\begin{minipage}[c]{63mm}
  $\mathrm{PVI}_i$, 
  $\mathrm{H3}_a^{\sigma}$,
  $\mathrm{H4}_a^{\sigma}$,  
  $\mathrm{H1}_i^{\sigma}$,
  $\mathrm{RelH1}_i$,
  $\mathrm{RelH3}_a$,
  $\mathrm{MFCC}_a(11)$,
  $\mathrm{MFCC}_i(6)$
  $\Delta\mathrm{MFCC}_a(1)$,  
  $\Delta\mathrm{MFCC}_a(3)$, 
  $\mathrm{GNE}_a^{\sigma}$, 
   \vspace{1mm}
\end{minipage}
  & \begin{minipage}{30mm}\centering\vspace{1mm} $Acc = 82.9\pm2.8$ \\ $Sens = 78.0\pm4.4$ \\ $Spec = 87.6\pm2.6$ \vspace{1mm} \end{minipage}  \\
 \hline
 \multicolumn{3}{|c|}{\footnotesize  Low order model}\\
 \hline
 \begin{minipage}[c]{16mm}\vspace{1mm} 10 best LASSO features +BSS ($N=5$) \vspace{1mm} \end{minipage}  &
 \begin{minipage}[c]{63mm}
  $\mathrm{PVI}_i$, 
  $\mathrm{MFCC}_i(2)$, 
  $\mathrm{MFCC}_i(9)$,
 $\mathrm{MFCC}_a(8)$, 
 $\mathrm{MFCC}_a(10)$,
 \vspace{2mm}
 \end{minipage}
 & \begin{minipage}{30mm}\centering\vspace{2mm}  $Acc = 89.0\pm2.5$ \\ $Sens = 87.5\pm2.9$ \\ $Spec = 90.4\pm3.3$ \vspace{2mm} \end{minipage}  \\
\hline
\end{tabular}
\label{tbl:features}
\end{table}

\begin{figure}[thb]
  \centering
  \includegraphics[width=1\textwidth]{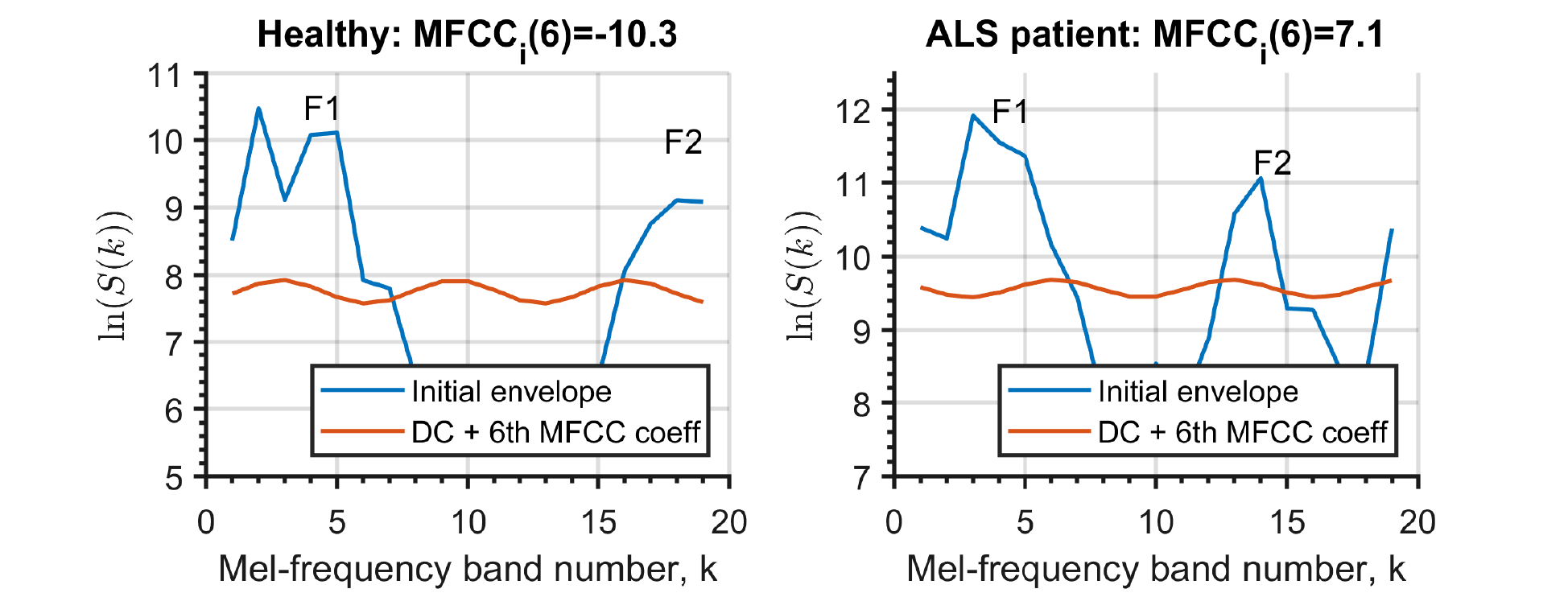}
  \caption{Interplay between features $F2_i$ and $\mathrm{MFCC}_i(6)$}
  \label{fig:mfcc_analysis}
\end{figure}

The analysis of tables~\ref{tbl:stat} and~\ref{tbl:features} leads to the logical question: why statistically significant feature $d_1$ was not selected by LASSO FS algorithm? Detailed analysis  have revealed that $d_1$ and $\mathrm{MFCC}_i(2)$ have strong correlation ($r=0.54$ with $p<1.0\cdot 10^{-5}$), 
thus, MFCCs already contain information possessed in $d_1$ feature. Another question: why such significant features like $F2_{conv}$ and $F2_i$ were not selected by neither algorithm? First of all these features are highly correlated ($r=0.85$ with $p<10^{-18}$), thus the location of $F2_i$ is more relevant rather than its proximity to $F2_a$. Furthermore, $F2_i$ is strongly correlated with $\mathrm{MFCC}_i(2)$ and $\mathrm{MFCC}_i(6)$ ($r=0.48$ and $r=-0.44$, accordingly), therefore the information about $F2_i$ can be passed to classifier with any of these parameters. LASSO and QoV algorithms have selected $\mathrm{MFCC}_i(4)$ that contained information about $F2_i$ location, while RelieFF algorithm selected $\mathrm{MFCC}_i(6)$ for this purpose. A visual example of  interplay between $F2_i$ and  $\mathrm{MFCC}_i(6)$ is given in figure~\ref{fig:mfcc_analysis}.   

Figure~\ref{fig:mfcc_analysis} shows the estimation of spectral envelopes computed during MFCC calculation and partial reconstruction of envelopes using DC component and 6-th MFCC coefficient. It can be seen that the voice of ALS patient is characterized by reduced frequency of second formant. As a result, projection onto 6th basis function of discrete cosine transform which is used in MFCC calculation is changing sign (if we compare HC and ALS voices). 

The result of our study confirm the findings of~\cite{Tsanas-2012}, where MFCC are also found to be highly informative features for Parkinson's disease detection. However unlike~\cite{Tsanas-2012} we give interpretation that MFCC reflect changes in second formant of vowel /i/ for ALS patients. Also it should be noted that proposed features extracted using harmonic analysis of the vowels are essential for obtaining good classifier. For example, among the 32 features selected by LASSO, ten describe the harmonic structure. Among the rest features: 9 MFCCs describe the envelopes of vowels (6 for /a/ and 3 for /i/), 7 delta MFCCs reflect variability of vowels envelopes, the GNE parameters gives information about noise content of the voice and PVI describes the changes in vibrato. Another interesting observation is that in subset of features selected by LASSO 19 are related to vowel /a/ and 13 to vowel /i/. It means that information contained in phonation /a/ is relevant and necessary for gaining high classification accuracy. It is interesting that traditional measures such as jitter, shimmer and HNR are out of table~~\ref{tbl:features}. This suggests that PVI, MFCC and harmonic structure parameters have greater predictive power for distinguishing between HC and patients with ALS.

Surely, that main goal of classification is most accurate detection of the ALS patients voices. In this regard, LDA model with 32 features and 99\% accuracy is a significant result. However, there is reason to believe that this feature set is quite specific for our voice database. More relevant information about parameters that are most important for ALS detection can be derived by analyzing high-performance LDA models with small number of features. Table~\ref{tbl:features} shows that LDA models obtained using the QoV and RelieFF feature selection algorithms have a small number of features, however they have quite low performance. To find a model with higher performance we took LDA classifier with 10 best feature picked out using LASSO algorithm, which has accuracy 87.6$\pm$2.6\% ($Sens$ = 90.5$\pm$4.1\%, $Spec$ = 84.8$\pm$3.0\%) and applied back-step selection procedure. As a result LDA model with five features ($\mathrm{MFCC}_i(2)$, $\mathrm{MFCC}_i(9)$, $\mathrm{PVI}_i$, $\mathrm{MFCC}_a(8)$, $\mathrm{MFCC}_a(10)$) was obtained, which has accuracy 89.0\% (see last row of table~~\ref{tbl:features}). Thus, it can be concluded that the most important information for detecting the ALS patients’ voices is contained in the spectral envelopes of sounds /a/ and /i/ (MFCC parameters), as well as in the vibrato changes (PVI).

\begin{table}[thb]\centering 
\footnotesize
\caption{Comparison with other studies} 	
\renewcommand{\arraystretch}{1.2} 
\begin{tabular}{|l|c|c|c|c|}
\hline \hline
 & Norel~\cite{Norel-18}  & 
   Spangler~\cite{Spangler-17} & 
   An~\cite{Kwanghoon-2018} & 
   Present \\
\hline
 
Feature & 
 \begin{minipage}{18mm}\vspace{2mm}\centering 
 Extracted with Open-SMILE toolkit \vspace{2mm}\end{minipage} & \begin{minipage}{18mm}\vspace{2mm}\centering 
 Fractal jitter, MFCC, RPDE + articulatory data \vspace{2mm}\end{minipage}& 
 \begin{minipage}{18mm}\vspace{2mm}\centering 
 filterbank energies + its deltas \vspace{2mm}\end{minipage} & 
 \begin{minipage}{18mm}\vspace{2mm}\centering 
 MFCC, Harmonic parameters, PVI \vspace{2mm}\end{minipage}\\
 
\begin{minipage}{18mm}  \vspace{2mm} 
Total number of features \vspace{2mm} \end{minipage}& 
\begin{minipage}{18mm}\vspace{2mm}\centering 
   \textit{for male} 1 \\  \textit{for female}~15 \vspace{2mm}\end{minipage} &
\begin{minipage}{18mm}\vspace{2mm}\centering 
  17 \vspace{2mm}\end{minipage}& 
\begin{minipage}{18mm}\vspace{2mm}\centering 
  120 \vspace{2mm}\end{minipage} & 
\begin{minipage}{18mm}\vspace{2mm}\centering 
  32 \vspace{2mm}\end{minipage}\\
 
Classifier & 
 \begin{minipage}{18mm}\vspace{2mm}\centering linear SVM \vspace{2mm}\end{minipage} & 
 \begin{minipage}{18mm}\vspace{2mm}\centering Extreme Gradient Boosting \vspace{2mm}\end{minipage}& CNN  & LDA\\
 
Verification &
\begin{minipage}{18mm}\vspace{2mm}\centering Leave-five-subject-out CV
 \vspace{2mm}\end{minipage} &
\begin{minipage}{18mm}\vspace{2mm}\centering Leave-one-subject-out CV
	\vspace{2mm}\end{minipage} & 
\begin{minipage}{18mm}\vspace{2mm}\centering Leave-one-subject-pair-out CV
	\vspace{2mm}\end{minipage} & 
\begin{minipage}{18mm}\vspace{2mm}\centering 8-fold CV
	\vspace{2mm}\end{minipage}\\

Database & 
\begin{minipage}{18mm}\vspace{2mm}\centering 133 speakers \\ (67 ALS, \\66 HC), \\ running speech
 \vspace{2mm}\end{minipage} & 
\begin{minipage}{18mm}\vspace{2mm}\centering 83 speakers \\ (49 ALS, \\34 HC), \\ DDK test
 	\vspace{2mm}\end{minipage} & 
\begin{minipage}{18mm}\vspace{2mm}\centering 26 speakers \\ (13 ALS, \\13 HC), \\ running speech
	\vspace{2mm}\end{minipage} & 
\begin{minipage}{18mm}\vspace{2mm}\centering 64 speakers \\ (31 ALS, \\33 HC), \\ SVP test
	\vspace{2mm}\end{minipage} \\

\begin{minipage}{18mm} \vspace{2mm} 
Reported performance \vspace{2mm}\end{minipage} & 
\begin{minipage}{18mm}\centering  \vspace{2mm} \textit{for male} \\Acc=79\%\\Sens=76\%\\ Spec=70\% \\ \textit{for female}
\\Acc=83\%\\Sens=78\%\\ Spec=88\% \vspace{2mm} \end{minipage} & 
\begin{minipage}{18mm}\centering \vspace{2mm} Acc=90.2\%\\ Sens=94.2\%\\ Spec=85.1\% \vspace{2mm} \end{minipage} & 
\begin{minipage}{18mm}\centering \vspace{2mm} Acc=76.2\%\\ Sens=71.5\%\\ Spec=80.9\% \vspace{2mm} \end{minipage} & 
\begin{minipage}{18mm}\centering \vspace{2mm} Acc=99.7\%\\ Sens=99.3\%\\ Spec=99.9\% \vspace{2mm} \end{minipage} \\
\hline
\end{tabular}
\label{tbl:compare}
\end{table}

Table~\ref{tbl:compare} presents comparison of the present work with recent similar studies. The purpose of those works was to discriminate between healthy people and ALS patients. The main differences between these studies concern speech tasks, classification approaches, features and verification methods.
The most closest result was obtained in~\cite{Spangler-17}. However, in~\cite{Spangler-17} along with voice recording articulatory data was used.  In table~\ref{tbl:compare} two different performance results are given for study~\cite{Norel-18} because it uses sex-specific features for classifiers to take into account differences in the vocal tracts of males and females. 
Study~\cite{Kwanghoon-2018} presents results of two type: sample-level and person level classification. The second type is obtained based on sample voting. In table~\ref{tbl:compare} we compare only sample-level classifiers. However, even person-level classifier based on 5 samples~\cite{Kwanghoon-2018} has accuracy 90.8\%, sensitivity 85.6\% and specificity 94.9\%. Therefore the obtained result with near 99\% of accuracy, sensitivity and specificity based on LDA classifier can be considered as an essential improvement over the previous results.

\subsection{Additional experiment: early ALS detection}
The following additional experiment has been performed in order to determine validity of LDA models with features extracted from SVP test for early ALS detection problem. From ALS patients were chosen 12 that having been diagnosed less than one year before recordings (see table~\ref{tbl:participant}). So, the reduced dataset included 45 speakers (33 HC + 12 ALS).

Using the reduced dataset, we performed feature selection procedures and optimization of feature set as described above. However, in contrast to experiments presented in previous sections, we used leave-one-subject-out (LOSO) cross-validation procedure to evaluate the performance of classifiers.~\cite{Flach-2012,hastie-01}. In fact, the LOSO method is a $k$-fold CV procedure, with $k$ equal to the size of the dataset. We used LOSO in order to bring closer the size of the samples on which the classifiers are trained in sections 4.2 and 4.3. In section 4.2, where the 8-fold CV was used, the LDA classifier model was trained on 54 samples, in this section, using the LOSO CV method, the classifier is trained on 44 samples.

\begin{table}[tbh]\centering 
	\caption{Early ALS detection: selected feature subsets and classification accuracy 
	} 	
	\renewcommand{\arraystretch}{1.2} 
	\begin{tabular}{|l|c|c|}
		\hline \hline
		\begin{minipage}[c]{18mm}\vspace{2mm} FS \\ algorithm \vspace{1mm}\end{minipage}   	& Features & Accuracy results (\%) \\
		\hline
		\hline
		\begin{minipage}[c]{18mm} QoV + BSS ($N=5$)
		\end{minipage}   	& 
		\begin{minipage}[c]{61mm}
			$\mathrm{H3}_a^{\sigma}$,
			$\mathrm{H5}_i^{\mu}$,
			$\mathrm{H6}_i^{\mu}$,
			$\mathrm{RelH6}_i$,			
			$\mathrm{MFCC}_i(6)$
		\end{minipage}
		& \begin{minipage}{30mm}\centering \vspace{2mm}  $Acc=84.4\pm5.4$ \\ $Sens=75.0\pm6.5$ \\ $Spec=87.9\pm4.9$ \vspace{2mm} \end{minipage}\\
		
		\begin{minipage}[c]{18mm} RelieFF ($N=5$) \end{minipage}  &
		\begin{minipage}[c]{61mm}
			$\mathrm{MFCC}_a(8)$,
			$\mathrm{MFCC}_a(11)$, 
			$\mathrm{MFCC}_i(2)$,
			$\mathrm{MFCC}_i(6)$,
			$\mathrm{PFR}_a$
			\vspace{2mm}
		\end{minipage}
		& \begin{minipage}{30mm}\centering\vspace{2mm}  $Acc = 93.3\pm3.7$ \\ $Sens = 83.3\pm5.6$ \\ $Spec = \mathbf{97.0\pm2.6}$ \vspace{2mm} \end{minipage}  \\
		
		\begin{minipage}[c]{18mm} LASSO ($N=12$) \end{minipage}  & 
		\begin{minipage}[c]{61mm}
			$d_1$,
			$\mathrm{PFR}_a$,
			$\mathrm{H7}_i^{\sigma}$,
			$\mathrm{RelH6}_a$,
			$\mathrm{MFCC}_a(6)$, 
			$\mathrm{MFCC}_a(8)$, 
			$\mathrm{MFCC}_i(2)$,
			$\mathrm{MFCC}_i(3)$,
			$\mathrm{MFCC}_i(6)$,
			$\mathrm{MFCC}_i(9)$,
			$\Delta\mathrm{MFCC}_i(6)$, 
			$\Delta\mathrm{MFCC}_i(12)$,

			\vspace{2mm}
		\end{minipage}
		& \begin{minipage}{30mm}\centering\vspace{2mm}  $Acc = \mathbf{95.6\pm3.1}$ \\ $Sens = \mathbf{91.7\pm4.1}$ \\ $Spec = \mathbf{97.0\pm2.6}$ \vspace{2mm} \end{minipage}  \\
		\hline
	\end{tabular}
	\label{tbl:early_als}
\end{table}

LDA model with 5 features and above 80\% accuracy has been obtained using QoV feature selection algorithm with BSS procedure (see table~\ref{tbl:early_als}). Best LDA model obtained using Relief algorithm with BSS procedure has 39 features and 100\% accuracy. The same accuracy is achieved by the LDA model using 28 features selected by LASSO algorithm. However, these feature sets (unless they legitimacy) are too specifically fit to our database. We believe that more relevant conclusions can be derived by analyzing models with feature sets of limited size. For example, LDA model trained on the first 5 features selected by RelieFF algorithm has 93,3\% accuracy (see table~\ref{tbl:early_als}). Furthermore, among the LDA models with a small number of features we can highlight one that has 95,6\% accuracy and trained on the first 12 features selected by the LASSO algorithm.

Analyzing the features contained in the table~\ref{tbl:early_als}, we can draw the following conclusions. In all feature sets $\mathrm{MFCC}_i(6)$ is present, its relevance is discussed in previous sections (for example, see figure~\ref{fig:mfcc_analysis}). Four out of five features selected by RelieFF algorithm are also included in feature set picked out by LASSO algorithm. This indicates their high significance for early ALS detection. Feature set obtained using the RelieFF algorithm shows that valuable information for early ALS detection is contained in spectral envelopes of the vowels /a/ and /i/ (this information is concentrated in parameters $\mathrm{MFCC}_a(8)$, $\mathrm{MFCC}_a(11)$ and $\mathrm{MFCC}_i(2)$, $\mathrm{MFCC}_i(6)$). Parameter $\mathrm{PFR}_a$, which indicates the degree of fundamental frequency variation, is also important for early ALS detection. It should be noted that neither of the feature sets contains parameters PVI and PPE, the significance of which was revealed in the previous experiment. This means that changes in the vibrato are not related to the early diagnosis of the ALS, but rather characteristic of later stages of the disease.

\section{Conclusion}
\label{sec:conclusion}
In this study we investigate the possibility of designing linear classifier for discriminate ALS patients from  healthy controls using acoustical sustained vowels /a/ and /i/ phonation tests.  A large set of features was analysed. 
LDA classifier with 99.7\% accuracy (99.3\% sensitivity, 99.9\% specificity) was obtained based on 32 features determined by LASSO feature selection algorithm. We also obtained the LDA model with only 5 features that has 89.0\% accuracy (87.5\% sensitivity, 90.4\% specificity). We found that the most important information for detecting the ALS patients’ voices is contained in the spectral envelopes of sounds /a/ and /i/ (MFCC parameters), as well as in the vibrato changes (PVI). Like in~\cite{Spangler-17} traditional jitter measures were found not to have a high importance. We also carried out experiment to determine validity of LDA models with features extracted from SVP test for early ALS detection problem. Our results show that it is possible to obtain LDA model with 93.3\% accuracy (83.3\% sensitivity, 97.0\% specificity) based on only 5 features determined by RelieFF algorithm. We can also draw the conclusion that valuable information for early ALS detection is contained in spectral envelopes of the vowels /a/ and /i/ (MFCC parameters). We also found that the selected feature sets did not contain the PVI and PPE parameters. This means that changes in the vibrato  are not related to the early diagnosis of the ALS, but rather characteristic of the later stages of the disease.
It should be noted that the data for this study was collected using smartphone with regular headset. Therefore we can assert that proposed approach is tolerant to non-professional recording condition. 

\section*{Acknowledgements}
The authors thank the anonymous reviewers for their useful comments.

\bibliography{mybibfile}

\end{document}